%% file: 190114C.tex
\documentclass[12pt]{article}

\usepackage{scicite}

\usepackage{epsfig,graphicx}
\usepackage{times}
\usepackage{ccaption}
\usepackage{upgreek}
\usepackage{amsmath,amssymb}
\usepackage{url}
\usepackage{hyperref}
\usepackage[hang,flushmargin]{footmisc} 
\usepackage{longtable} 
\usepackage{booktabs}
\usepackage{deluxetable}

\newenvironment{sciabstract}{%
\begin{quote} \bf}
{\end{quote}}

\title{GRB 190114C: An Upgraded Legend} 
\author
{Yu Wang$^{1,2}$, Liang Li $^1$, Rahim Moradi $^{1,2}$, Remo Ruffini $^{1,2,3,4,5,6}$\\
\\}
\date{}

\begin{document} 
\baselineskip24pt

\maketitle 

\noindent
\normalsize{$^{1}$ICRANet, P.zza della Repubblica 10, 65122 Pescara, Italy.}\\
\normalsize{$^{2}$ICRA and Dipartimento di Fisica, Sapienza Universit\`a di Roma, P.le Aldo Moro 5, 00185 Rome, Italy.}\\
\normalsize{$^{3}$ICRANet - INAF, Viale del Parco Mellini 84, 00136 Rome, Italy.}\\
\normalsize{$^{4}$Universit\'e de Nice Sophia Antipolis, CEDEX 2, Grand Ch\^{a}teau Parc Valrose, Nice, France.}\\
\normalsize{$^{5}$ICRANet-Rio, Centro Brasileiro de Pesquisas F\'isicas, Rua Dr. Xavier Sigaud 150, 22290--180 Rio de Janeiro, Brazil.}\\
\normalsize{$^{6}$ICRA, University Campus Bio-Medico of Rome, Via Alvaro del Portillo 21, I-00128 Rome, Italy.}\\

{\let\thefootnote\relax\footnote{
\normalsize{yu.wang@uniroma1.it}, \\  \normalsize{liang.li@icranet.org}, \\ \normalsize{rahim.moradi@icranet.org}, \\
\normalsize{ruffini@icra.it}
}}
\pagebreak

\baselineskip24pt 

\begin{sciabstract}
Gamma-ray burst (GRB) 190114C first resembles the legendary GRB 130427A: Both are strong sources of GeV emission, exhibiting consistent GeV spectral evolution, and almost identical in detail for the morphology of light-curves in X-ray, gamma-ray and GeV bands, inferring a standard system with different scales. GRB 190114C is richer than GRB 130427A: a large percentage of $\sim 30\%$  energy is thermal presenting in the gamma-ray prompt emission, making it as one of the most thermal-prominent GRBs; Moreover, GRB 190114C extends the horizon of GRB research, that for the first time the ultra-high energy TeV emission ($> 300$~GeV) is detected in a GRB as reported by the MAGIC team. Furthermore, GRB 190114C urges us to revisit the traditional theoretical framework, since most of the GRB's energy may emit in the GeV and TeV range, not in the conventional MeV range.  Since GRB 190114C confidently supports that MeV and GeV emissions have the same origin, it helps to establish a new comprehensive acceleration and radiation mechanism. Overall, GRB 190114C refreshes our GRB knowledge and challenges our current GRB theoretical interpretation. 
\end{sciabstract}

In the past 50 years, gamma-ray burst (GRB) has been one of  the main focuses of astrophysical research. Its phenomena have been complemented gradually by two paths. The first path is the new generations of satellites expanding the observational capacity: \emph{Vela} detected the first GRB on July 2, 1967; \emph{BeppoSAX} confirmed GRB's cosmological origin with the first redshift measurement was made for GRB 970508 (z=0.835) \cite{1997Natur.387..878M} and discovered the first X-ray \cite{1997Natur.387..783C} and the optical \cite{1997Natur.386..686V} afterglow; \emph{Swift} \cite{2004ApJ...611.1005G} broadened the observations in number and in great detail; \emph{Fermi} \cite{2009ApJ...702..791M} made possible to analyse the GeV spectrum. The other path relates to specific GRBs carrying rich or/and characteristic information: GRB 980425 revealed the GRB and supernova association \cite{1998Natur.395..670G, 1998Natur.395..663K}; GRB 090902B exhibits thermal domination \cite{2010ApJ...709L.172R}; GRB 130427A brought the unprecedentedly abundant and long-lasting GeV observation \cite{2014Sci...343...48M, 2014Sci...343...42A,2014Sci...343...38V,2014Sci...343...51P}, GRB 130603B displayed the kilonova signal \cite{2013Natur.500..547T}; GRB 170817A coincided with the detection of gravitational wave \cite{2017ApJ...848L..13A}. The current GRB 190114C combines both paths. First, MAGIC telescope highlights the fast follow-up observation of GRBs in its science program since its inception, and TeV photon was eventually found in GRB 190114C. Second, GRB 190114C epitomises the specific GRBs features, that the strong and long duration GeV emission, the high thermal abundance, and the representative spectral evolution, together with its characteristic TeV emission, GRB 190114C presents to us the most complete portrait of a GRB hitherto. 

In the meanwhile, the theoretical models have been developed, and some concepts are widely accepted. For instance, the black hole or the pulsar acts as the central engine \cite{2018ApJS..236...26L}; the synchrotron or/and the Compton up-scattering generate the non-thermal emission, the photosphere generates thermal emission \cite{2005ApJ...625L..95R,2009ApJ...702.1211R}, these emissions produce the majority energy of a GRB in the photons energy band of tens of keV to several MeV. The spectral energy distribution of GRB 190114C may challenge some of these concepts, it elevates the dominating energy band to more than $10$ MeV, our first analysis proposes that the spectra of energy lower and higher than $10$ MeV are correlated, but can hardly be modelled by a simple combination of synchrotron and Compton up-scattering processes. In this article, we sail from the legendary GRB 130427A, which presents almost identical temporal evolution as GRB 190114C. We then explore the traits of GRB 190114C, of its significant thermal emission and high energy emission. From the findings, GRB 190114C alters the focus of radiative modelling to higher energy, our analysis first finds a strong correlations of gamma-ray and GeV emissions, it infers the same origin, discussions and proposes company.

\section*{Observation} 
\label{sec:observation}


At 20:57:02.63 UT on 14 January 2019, gamma-ray burst (GRB) 190114C triggered the Gamma-ray Burst Monitor (GBM) onboard the \emph{Fermi} Gamma-ray Space Telescope, here we take this trigger time as $t_0$, and define $T90=116$~s by the \emph{Fermi}-GBM. The Swift Burst Alert Telescope (BAT) was triggered as well, it slewed immediately to the burst and localised the position to R.A. = $54.5068^\circ$, Dec. = $-26.9467^\circ$ (J2000) with an uncertainty of 5 arcsec \cite{GCN23688}, \emph{Swift}'s X-Ray Telescope (XRT) began observing the field $64$~s after the trigger. The hard X-ray satellite Insight-HXMT was triggered simultaneously, it observed a light curve of multiple pulses \cite{2019GCN.23716....1X}. Various optical telescopes followed the observation: MASTER \cite{GCN23690,GCN23693}, Pan-STARRS \cite{GCN23692}, NOT \cite{GCN23695}, OASDG \cite{GCN23699}, and GROND  \cite{GCN23702}.  This is one of the largest collaborations, of more than $30$ satellites and telescopes joined the observation. The redshift of this GRB is detected as $z=0.42$ by NOT \cite{GCN23695} and confirmed by GTC \cite{GCN23708}. At the trigger time,  the bore-sight of \emph{Fermi}'s Large Area Telescope (LAT) was $67^\circ$ from the GRB, this bore-sight angle keeps increasing, at time $\sim 200$~s, it passes the threshold of $75^\circ$ degree, then re-enters at the time later than $\sim 8000$~s when the GeV emission is supposed to be weak \cite{GCN23709}. The MAGIC team reported that their data show a clear excess of gamma-rays more than $300$~GeV with significance $>20$ sigma from $50$~s to $1200$~s \cite{GCN23701}. Therefore, there exists a common time interval from $\sim 50$~s to $\sim 200$~s covered by \emph{Fermi} and MAGIC, and partially covered by \emph{Swift}, MASTER and other lower energy telescopes.


\section*{Sail from GRB 130427A to GRB 190114C} 
\label{sec:twin_grbs}

\begin{figure*}[ht]
\centering
\includegraphics[width=1.0\hsize,clip]{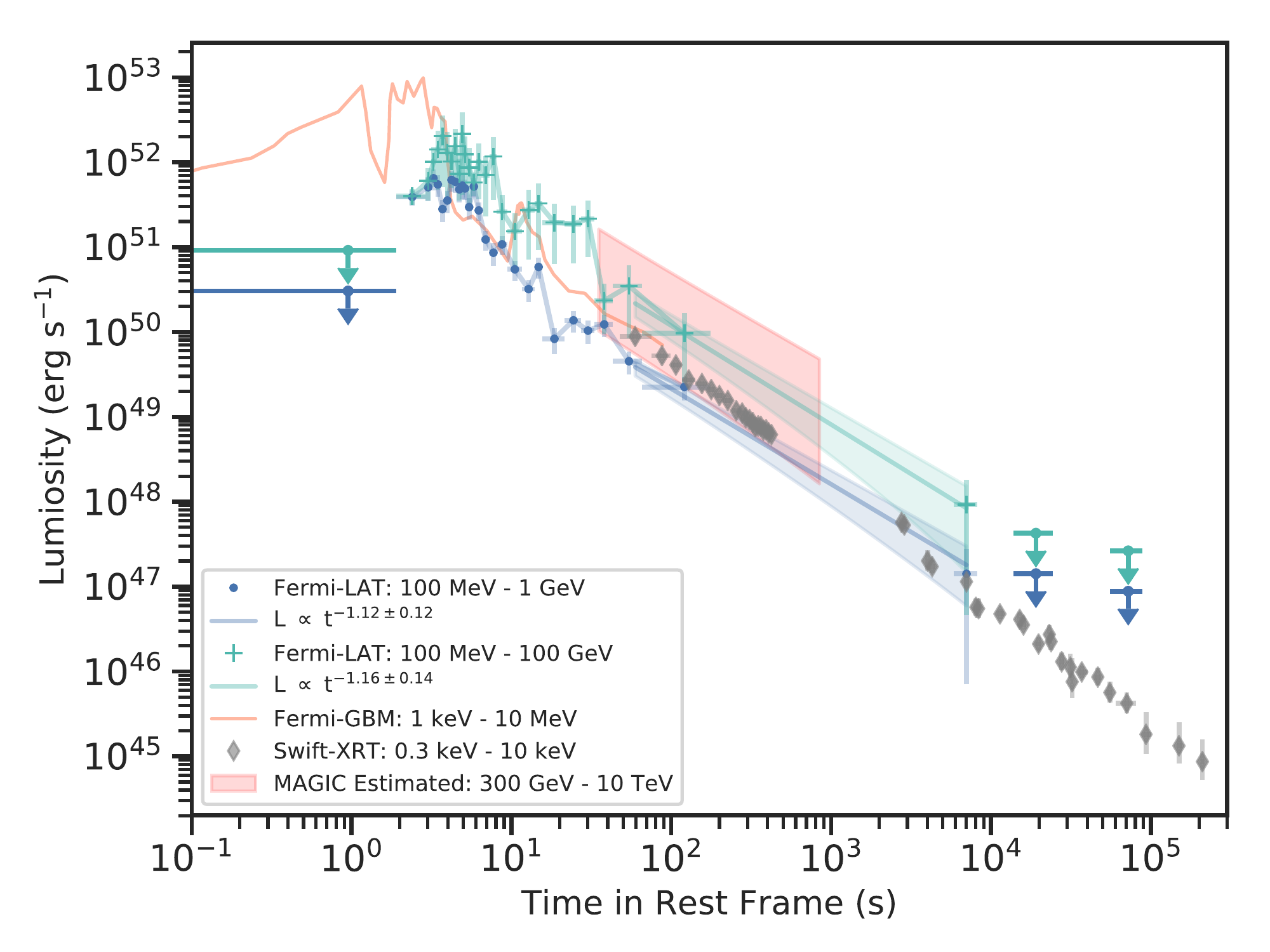}
\caption{Luminosity light-curve of \textit{Fermi}-GBM, \textit{Fermi}-LAT, \textit{Swift}-XRT and the estimation of the MAGIC light-curve. Blue and green colours represent the energy of $100$~MeV to $1$~GeV and $100$~MeV to $100$~GeV observed by Fermi-LAT, orange colour is the Fermi-GBM observation extrapolated to $1$~keV to $10$~MeV, grey colour is for the soft X-ray from $0.3$~keV to $10$~keV observed by Swift-XRT. The Fermi-LAT light-curves are fitted by a power-law using the data points $20$~s after the trigger time, where the prompt pulses end. The red region is the estimation of the MAGIC observation at energy $300$~GeV to $10$~TeV from $50$~s to $1200$~s, a TeV bump may exist before $100$~s.}
\label{fig:190114C_luminosity}
\end{figure*}

The structure of the gamma-ray light-curve observed by \emph{Fermi}-GBM includes two major pulses. The first bright pulse has multi-peaks, it starts with a peak beginning at the trigger time, lasting $2.3$~s, followed by a group of complex peaks till $6$~s. The second dim pulse starts at $\sim 15$~s, showing a fast rise/exponential decay (FRED) behaviour till $\sim 25$~s. The above description of the prompt emission structure can be safely applied on the second and the third pulses of GRB 130427A with the modified value. GRB 130427A has an initial $\sim 3$~s small pulse \cite{2014Sci...343...51P}, of which energy equals to $2\%$ of the energy in the second pulse. Such a relatively weak and short duration pulse is missed in the observation of GRB 190114C, but its existence cannot be confidently excluded, since the restriction of the telescopes capacity is possible to cause to the missing observation. \emph{Fermi}-GBM observation of GRB 130427A is saturated after the first pulse, we are unable to compare the gamma-ray spectral evolution of these two bursts.

\begin{figure}[ht]
\centering
\includegraphics[width=1\hsize,clip]{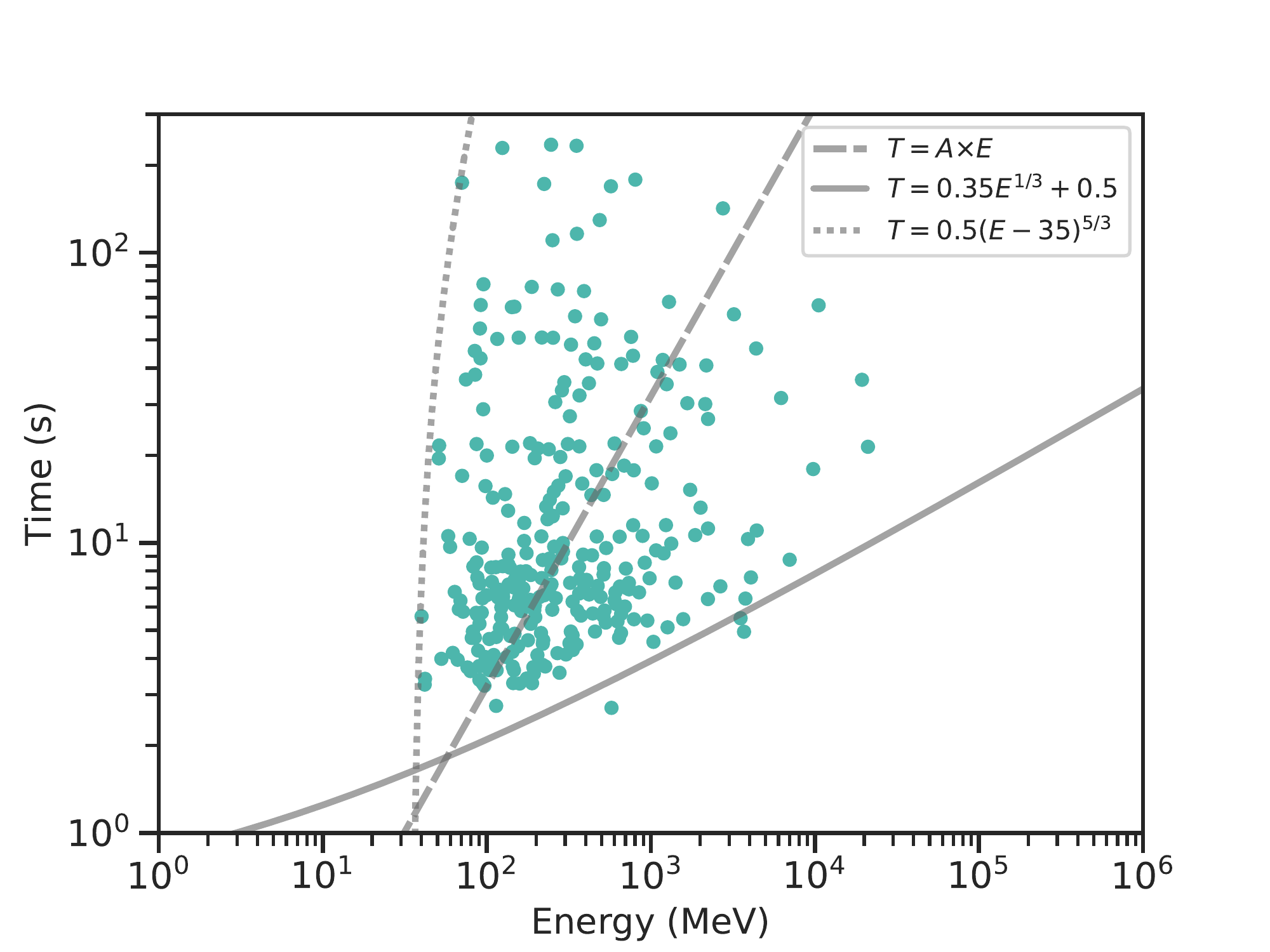}
\caption{Time delay of high energy photons. \emph{Fermi}-LAT photons associated with GRB 190114C with probability more than $50\%$ before $300$~s are plotted as green points. The solid line and the dotted line are fittings of the lower and upper boundaries of the data points; dashed line corresponds to the linear relation of LIV time delay. In the legend, $T$ is the photons arriving time and $E$ presents the photon energy, $A = 0.032$ is obtained from the LIV constrain $E_{\rm QG,1} = 0.63\times10^{16} ~ \rm GeV$ inferred from the time lag of Fermi-LAT spectra. The solid line of the lower boundary may present the initial photons, by extrapolating the solid line, a $\sim 100$~GeV photon is supposed to arrive at $\sim 20$~s, this coincides with the highest energy photon in the twin GRB 130427A \cite{2014Sci...343...42A}, and the first $1$~TeV photon may arrive at $\sim 35$~s, which could be confirmed by the releasing of MAGIC data.}
\label{fig:190114C_LIV}
\end{figure}

GRB 190114C is luminous in the GeV emission, its Test Statistics (TS) value of the $T90$ is $2520$, which is among the \emph{Fermi} GRBs with the highest TS value. More than $200$~photons with energy $>100$~MeV are observed, this number is similar to GRB 090902B and 090926A, the record holder is GRB 130427A, of which the number is $\sim 500$ \cite{2009ApJ...706L.138A,2010ApJ...718L..14S,2014Sci...343...42A}. The first  $>100$~MeV photon received with probability $>80\%$ is at $2.70$~s. The highest energy photon of $21.05$~GeV arrives at $21.43$~s, besides there are two more photons with energy more than $10$~GeV, they are of energy $19.36$~GeV and $10.54$~GeV, arriving at $36.47$~s and $65.86$~s respectively, figure \ref{fig:190114C_LIV} displays the arriving time of  the \emph{Fermi}-LAT photons. During the prompt emission, the GeV spectral evolution of GRB 190114C and GRB 130427A are consistent, that the spectra of all the time bins are best fitted by a single power-law, starting softly with a photon index $\sim -3$, then fluctuates between $\sim -2.5$ and $\sim -1.5$ \cite{2014Sci...343...42A}. The GeV luminosity light-curves of GRB 190114C and GRB 130427A are consistent as well, that they both have a delayed onset with respect to the gamma-ray emission, and they are dim during the strong gamma-ray emission, then become intense and reach the peak at the time when the gamma-ray emission has faded. After the early spiky structures, the GeV light-curve of 190114C follows a power-law decay with a power-law index $-1.16\pm0.14$, this is again consistent with GRB 130427A, of which the extended GeV light-curve has a power-law decay with index $-1.17\pm0.06$.

The soft X-ray light-curves of GRB 190114C observed by \emph{Swift}-XRT exhibits power-law decay behaviour, the power-law index is $-1.36\pm0.002$. The corresponding light-curve of GRB 130427A has an upwarp at the beginning, then decays as a power-law with index  $-1.31\pm0.01$. The early upwarp in GRB 130427A is possibly associated with its last prompt pulse. The \emph{Swift}-XRT observation of 190114C starts later than its last prompt pulse, therefore the existence of such an upwarp in GRB 190114C is unknown. The spectra are both best fitted by a single power-law, with photon index $-1.74\pm0.07$ for GRB 190114C, and $-1.70\pm0.08$ for GRB 130427A. 

\begin{figure}[ht]
\centering
\includegraphics[width=0.85\hsize,clip]{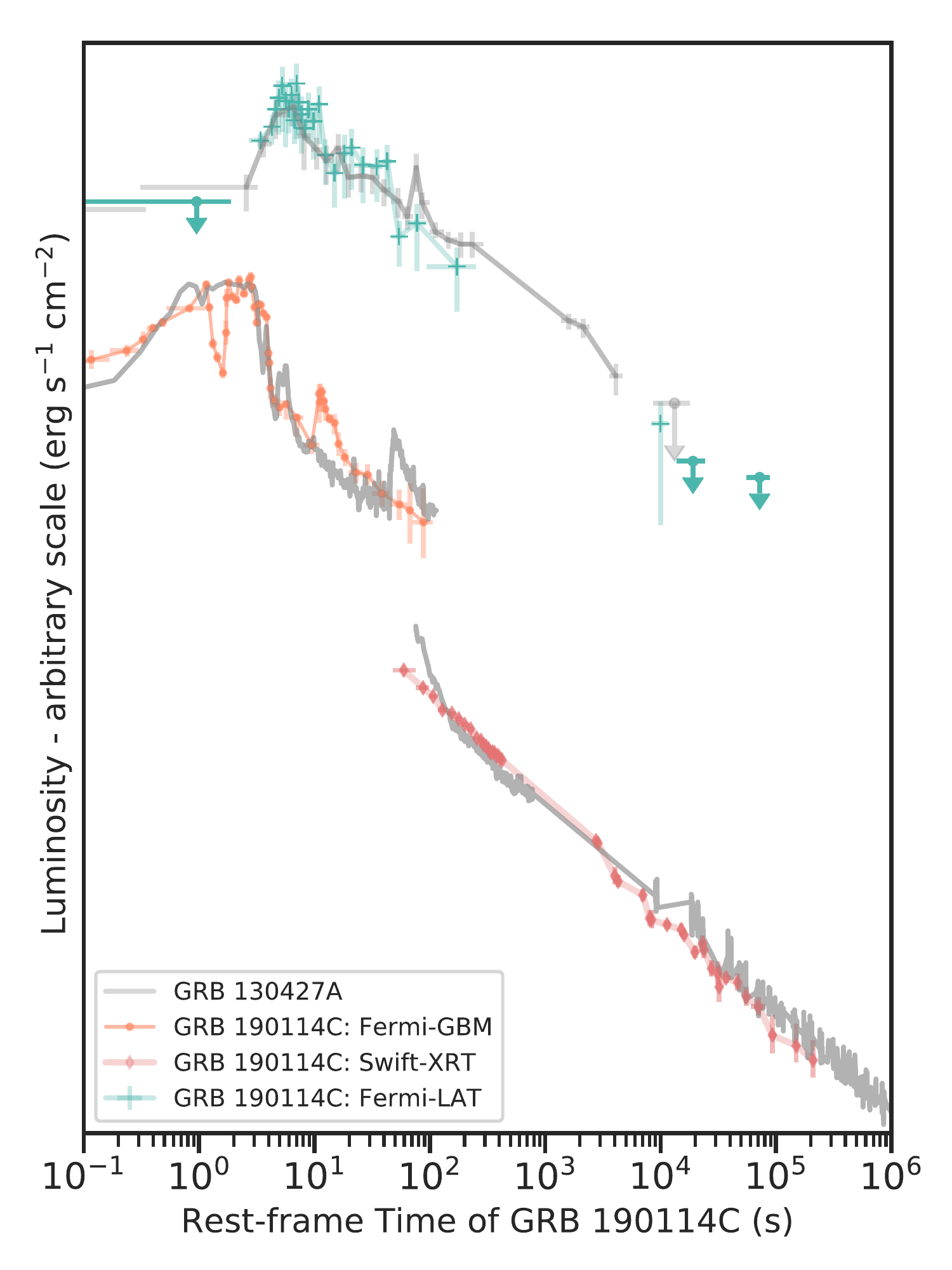}
\caption{Similar evolution of GRB 190114C and 130427A. The x-axis is the rest-frame time of GRB 190114C, the time of X-ray, gamma-ray and GeV light-curves of GRB 130427A $t$ are scaled by a linear transformation $0.53 \times (t-2.23)$~s. The y-axis is the luminosity, to have the overlap, we scale the X-ray, gamma-ray and GeV light-curves of 130427A by multiplying $1, 0.2$ and $ 10$, respectively. Then we arbitrarily shift the three groups of light-curve to have a clear view. These two GRBs generally have the same behaviour after the rescaling, and they infer the same system configuration with different scales. For comparing the unscaled light-curves, we refer to figure \ref{fig:190114C_luminosity} in this article and figure 8 in \cite{2015ApJ...798...10R}.}
\label{fig:190114C_index}
\end{figure}

GRB 190114C and GRB 130427A already look very similar from the above comparison, but this is not the whole story. We apply a linear transformation on the time coordinate of 130427A, $t' = 0.53 \times (t-2.23)$~s, where $t$ is the original time coordinate of GRB 130427A in the cosmological rest-frame, and $t'$ is the new transformed time coordinate, and rescale the luminosity of GRB 130427A by multiplying $0.2, 10$ and $1$ for gamma-ray, GeV, and X-ray, respectively. The shocking feature appears, that all the variational shapes of gamma-ray, GeV and X-ray light-curves of these two GRBs overlap respectively. This is the first time discovering two GRBs showing twin temporal evolution in all the energy bands. The simple linear transformation indicates that topologically GRB 190114C and GRB 130427A are identical. The simplest but reasonable explanation for the time-scale factor of $0.53$ relates to the size of the GRB system, GRB 190114C is more compact. The offset $-2.23$~s in the rest frame ($-3$~s in the observer's frame) is due to the aforementioned additional initial pulse in GRB 130427A. The difference in luminosities could be the consequence of different system sizes, there are many apparent reasons to be investigated, for instance, for different energies, there may exist different energy partitions from the central engine, different opacities in the burst region, different angular distributions of emissions, and different viewing angle from the satellite. This topologically identical feature infers, at least for some luminous GRBs, the existence of a standard GRB template, which we have been searched for years. Such a template will play important roles in the cosmology, for instance, the potential of being a standard candle laying from the current universe to the early dark age.

The existence of a supernova is expected in GRB 190114C since its twin GRB 130427A is associated with SN 2013cq\cite{Xu:2013f70,Wang:201968a}, this expectation is also proposed in \cite{Ruffini2019} from model-based reasoning. The observation has difficulties since the strong absorption along the line of sight \cite{Kann2019}, but with the effort of 50 days observational campaign, an associated supernova is eventually confirmed \cite{GCN23983}.

\section*{Exploring the traits of GRB 190114C} 
\label{sec:explore_traits_of_grb_190114c}

\begin{figure}[ht]
\centering
\includegraphics[width=1\hsize,clip]{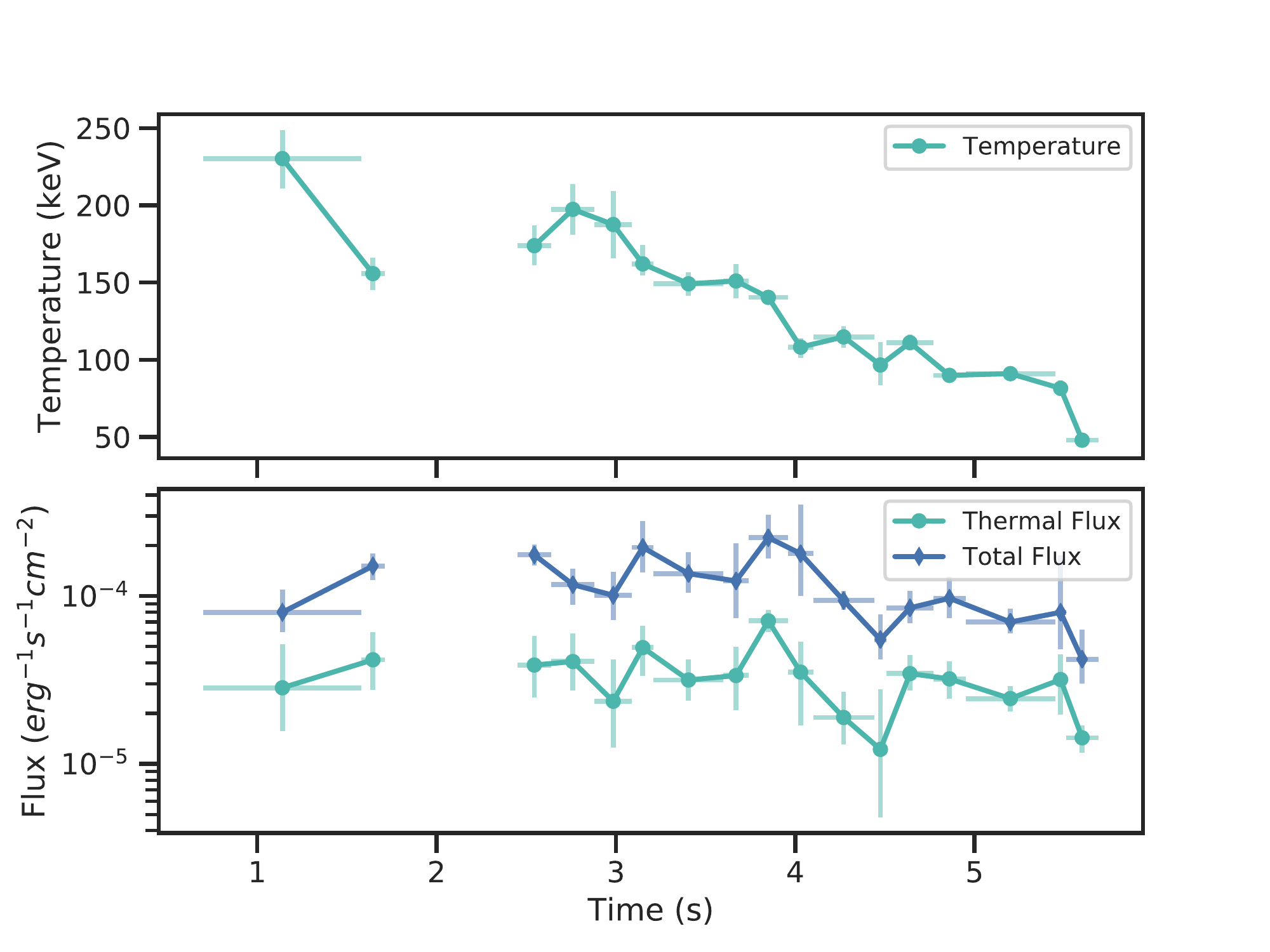}
\caption{The thermal evolution in the prompt emission. Upper:  The temperature evolution. Lower: The total flux by blue diamonds and thermal flux by green points, both are in the energy range of $1$~keV to $10$~MeV. The confident thermal component is found in the peak of the prompt pulse. The temperature drops with time, from $\sim 200$ keV to $\sim 50$~keV. The ratio of thermal flux is almost a constant of $\sim30\%$ within the error range. }
\label{fig:190114C_thermal}
\end{figure}

GRB 190114C contains more opulent properties than GRB 130427A. Its bright gamma-ray prompt emission observed by \emph{Fermi}-GBM makes the detailed time-resolved analysis possible. A time-resolved fitting of $T90$ using a cut-off power-law model shows that the low energy index $\alpha$ crosses the synchrotron limit of $-2/3$ \cite{1998ApJ...506L..23P} during the peaks of pulses. The spectra display a hard-to-soft evolution in the post-pulse phase, eventually $\alpha$ drops to be softer than the fast cooling limit of $-3/2$ \cite{1998ApJ...497L..17S}. The violation of the non-thermal synchrotron model encourages to search for the additional thermal component. Indeed, the nested models of cutoff power-law plus blackbody or band function plus blackbody are preferred for time bins within $0.70$~s to $1.71$~s and $2.45$~s to $5.69$~s, covering the peaks of the first pulse, there is no confident thermal in the second pulse at $\sim 10$~s. The temperature drops from $\sim 200$~keV to $50$~keV, while the ratio of the thermal flux keeps almost a constant $\sim 30\%$.  A time-integrated fitting of the $T90$ shows the consistency, $31\%$ of the \emph{Fermi}-GBM gamma-ray energy is in thermal, and the average temperature is $\sim 132$~keV. This high percentage of thermal flux reminds GRB 090902B, which is famous for the intense thermal emission appearing in the first half of the prompt emission \cite{2011MNRAS.415.3693R}. 

GRB 190114C is the first GRB confirmed the observation of TeV photons by MAGIC telescope. From the \emph{Fermi}-LAT light-curve, we may infer the intrinsic light-curve for the energy range of MAGIC during its operating time, from $50$~s to $1200$~s. We fit the Fermi-LAT spectrum of the entire MAGIC operating time, a power-law with photon index $-1.87\pm0.18$ is obtained. Assuming this power-law extends from GeV to TeV regime, we get the ratio of luminosity between the Fermi-LAT and MAGIC, that the MAGIC luminosity ($300$~GeV to $10$~TeV) is $0.369$ to $3.13$ times of the Fermi-LAT luminosity ($100$~MeV to $100$~GeV), then we shift the light-curve of Fermi-LAT to obtain the light-curve of MAGIC, plotted as the red region in figure  \ref{fig:190114C_luminosity}. This extrapolation of luminosity light-curve for MAGIC is here performed in the rest-frame of the GRB, the observed flux will be affectedly by the intervening processes during the propagation, e.g., the collision with the extragalactic background light may lower the observed flux two orders of magnitude for $1$~TeV \cite{2011MNRAS.410.2556D} photon at redshift $z \sim 0.4$.

\section*{Glance at a comprehensive model} 
\label{sec:glance_the_new_framework_of_grb}

GRB 190114C may change the definition of GRB by energy. Different from prompt gamma-ray emission, the GeV emission continues for a long time, at least till $8000$~s in GRB 190114C. To obtain the total energy of the GeV photons, it is essential to include the energy in the GeV afterglow. The resultant energy from \emph{Fermi}-LAT in the energy range of $100$~MeV to $100$~GeV is $1.8^{+1.3}_{-1.0} \times 10^{53}$~erg, equivalent to $\sim 75\%$ of its gamma-ray isotropic energy, which previously was considered as majority of a GRB's energy, of $2.47\pm0.22 \times 10^{53}$~erg released in the energy range of $1$~keV to $10$~MeV, this percentage is higher than other bright Fermi-LAT GRBs, of which $\sim 20\%$ \cite{2011ApJ...732...29C,2013ApJS..209...11A,2014Sci...343...42A}. From the spectrum of $10$~MeV to $30$~MeV observed by the BGO detectors on-boarded \emph{Fermi}-GBM, and the existence of $> 300$~GeV photons announced by MAGIC, it is reasonable to extrapolate the spectrum obtained from \emph{Fermi}-LAT to a wider energy range of $10$~MeV to $1$~TeV, within which the total energy is more than $3 \times 10^{53}$~erg, this value is higher than the traditional isotropic energy. Energy characterises the astrophysical phenomenon, the leading emission at high energy implies the procedure of forming a black hole is more extreme than our previous understanding. The change of the dominating energy band alters our focus of current GRB models, we are required to revisit the acceleration and the radiation processes.

\begin{figure*}
    \centering
    \includegraphics[width=1\hsize,clip]{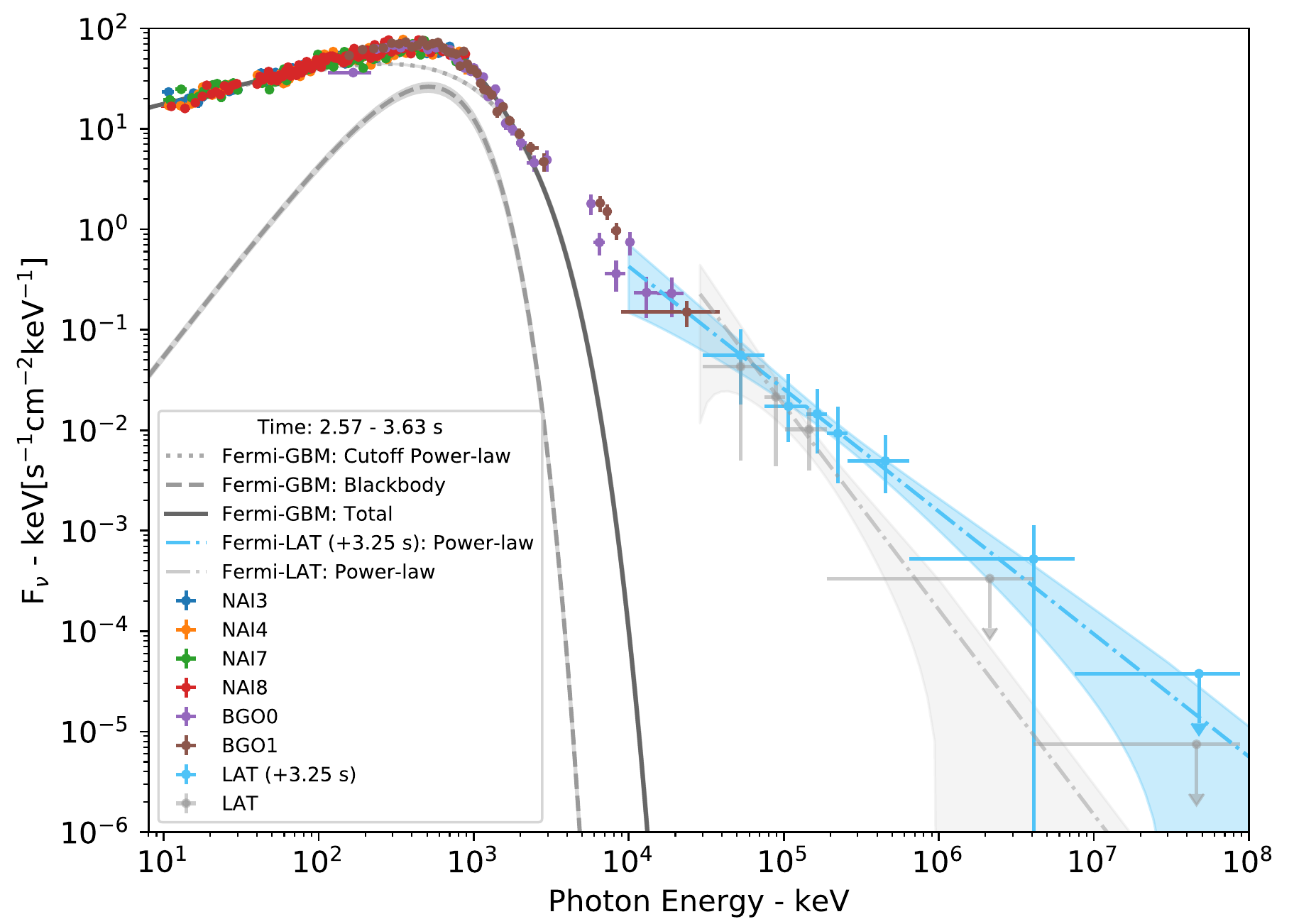}
    \caption{Multi-wavelength spectrum of $2.57$~s to $3.63$~s. The spectrum includes data from \emph{Fermi}-GBM (4 NAI and 2 BGO detectors) and \emph{Fermi}-LAT. The fitting of \emph{Fermi}-GBM is presented by a solid line, including the components of a blackbody function by a dashed line and a cutoff power-law by a dotted line. The \emph{Fermi}-LAT data plotted by light blue crosses are fitted by a power-law function as a dash-dot line. The fittings are performed separately, the Bayesian Monte Carlo iteration is used to fit the \emph{Fermi}-GBM data, and the unbinned maximum likelihood method, considering the background sources, is used to fit the \emph{Fermi}-LAT data. The extrapolation of \emph{Fermi}-LAT fitting connects the \emph{Fermi}-GBM data.}
\label{190114C_spectrum_gbm_257_363_lat_583_688_XSPEC}
\end{figure*}

\begin{figure}[ht]
\centering
\includegraphics[width=0.78\hsize,clip]{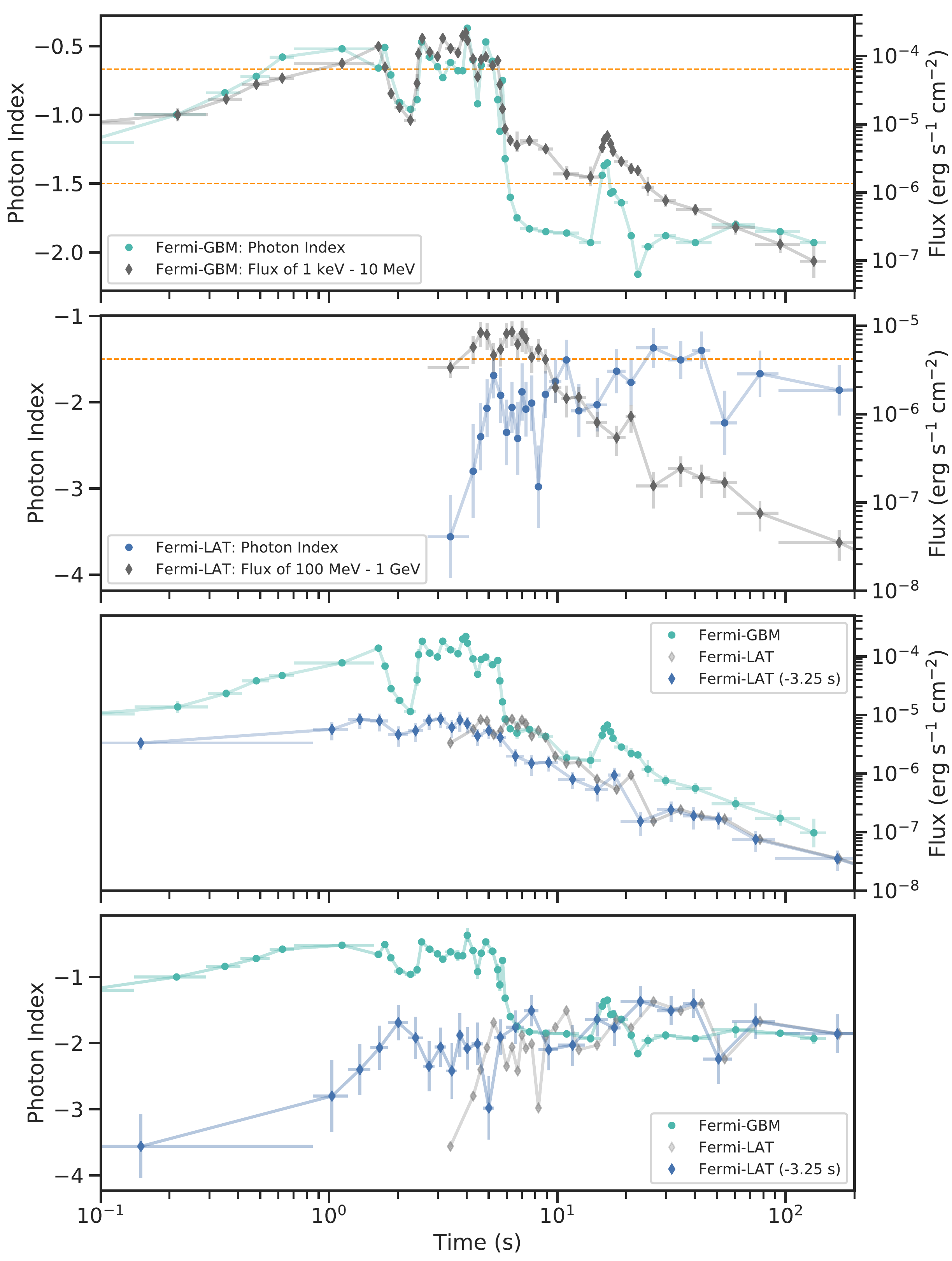}
\caption{Temporal evolution of flux and spectrum. From top to bottom are: (a) Observation from \emph{Fermi}-GBM. The low photon index (green dots) tracks The gamma-ray flux (grey diamonds). (b) Observation from \emph{Fermi}-LAT. The GeV photon index (blue dots) anti-tracks the GeV flux (grey diamonds). (c) The flux of gamma-ray obtained from \emph{Fermi}-GBM data in green dots, the flux of GeV obtained from \emph{Fermi}-LAT data in grey diamonds, and the GeV flux shifted $-3.25$~s in blue diamonds, which correlates with the gamma-ray flux. The correlation is clearer after normalising the flux, shown in the supplement figure \ref{fig:190114C_flux_norm_LAT_GBM}. (d) The photon index of gamma-ray fitted from \emph{Fermi}-GBM observation in green dots, the photon index of GeV fitted from \emph{Fermi}-LAT observation in grey diamonds, and the GeV photon index shifted $-3.25$~s in blue diamonds, which anti-correlates with the gamma-ray photon index. In these figures, \emph{Fermi}-LAT is limited to $100$~MeV to $1$~GeV, which contains most of the \emph{Fermi}-LAT photons, this energy range is more precise for the spectral fitting, it meets our purpose of having a clear view of correlations.}
\label{fig:190114C_LAT_GBM_Evolution}
\end{figure}

The prompt emission phenomenologically is synthesised by three components \cite{2011ApJ...730..141Z,zhang_2018}, most of the GRBs only have a dominating non-thermal component, some GRBs show a quasi-thermal component, and a few GRBs contain another non-thermal component extending to the high energy. GRB 190114C integrates all three components distinctly, its first pulse is quasi-thermal, its second pulse is non-thermal, and the GeV emission presents in both pulses. Previously, thermal and non-thermal emissions were thought to be correlated \cite{2014ApJ...784L..43B}, the high energy GeV emission was considered to evolve independently \cite{2011ApJ...729..114A,2014Sci...343...42A}. The high-resolution time-resolved analysis of the gamma-ray and GeV spectra in 190114C evidence all the three components are correlated. To find the correlation, we need to revisit the way of performing multi-wavelength data analysis, to be guided by the physical process of radiation. Instead of fitting all bands data that received at the same moment, we are supposed to fit the data from the same origin or associated origins. From the observation, there is no apriori knowledge to correlate some gamma-ray photons and some GeV photons, therefore, we first fit the data separately, then to look for their correlations. To our surprise for the simplicity, the cross-correlation method \cite{1988ApJ...333..646E} indicates a statistically strong correlation between the GBM and the LAT flux evolution. For a time displacement of $3.25$~s, the correlation coefficient reaches a peak value of $0.917$. This time displacement conforms to the delayed time of the first \emph{Fermi}-LAT photon arriving at $\sim 2.7$~s, the difference of $0.55$~s may be account for by the missed initial pulse as we discussed in section \ref{sec:twin_grbs}. In other words, after aligning the starting time of gamma-ray and GeV emissions, a distinct correlation of flux emerges. In addition, after shifting $3.25$~s accordingly for the spectral evolution, the power-law index ($\alpha_{GeV}$) of the GeV spectra anti-correlates with the low index of the gamma-ray spectrum ($\alpha_{\gamma}$), with a correlation coefficient $-0.631$, and it indicates $\alpha_{GeV} + \alpha_{\gamma} = 3\sim4$.  These correlations are consistent with the flux-tracking behaviour, that for gamma-ray emission, the lower photon index ($\alpha_{\gamma}$) tracks the gamma-ray flux, and for the GeV emission, the photon index ($\alpha_{GeV}$) anti-tracks the GeV flux. All the correlations are shown in figure \ref{fig:190114C_LAT_GBM_Evolution}, they can be confirmed visually as well, for instance, the spikes of the flux curve at $\sim 1.5$~s, $\sim 5$~s and $\sim 15$~s all have counterparts. And the spectral anti-correlation also has a spike to spike correspondence during the first pulse ($< \sim 7$~s). A more precise correlation possibly can be achieved by a more intricate time transformation, the time displacement can be a function of the time itself, since the GRB system is dynamic. From the flux evolution, a possible trend indicates the time displacement grows with time. An example of the multi-wavelength spectrum is given in figure \ref{190114C_spectrum_gbm_257_363_lat_583_688_XSPEC}, clearly the time-shifted \emph{Fermi}-LAT spectrum has connection to the \emph{Fermi}-GBM spectrum, that the extrapolation of the \emph{Fermi}-LAT spectrum coincides with the excess of \emph{Fermi}-GBM spectrum at $>3$~MeV. This feature generally exists in the spectra of other time bins and in the time-integrated spectrum of $T90$, it confirms that gamma-ray and GeV emissions are correlated.



GRB 190114C alters the focus of GRB emission to the high energy, our new philosophy of data analysis discovers the correlation of gamma-ray and GeV emissions, a new comprehensive modelling of multi-wavelength emission is urged. It shall be capable to interpret: 1) higher energy photon has a longer decay of arriving time; 2) gamma-ray and GeV are related, that their fluxes are positively correlated, and spectral indices are negatively correlated; 3) spectrum has a hard photon index $\sim -0.5$ at sub-MeV, and an additional component described by a Planck's function; 4) production of TeV photons. In the following, we attempt to take a glance at the comprehensive model.

The gamma-ray emission during the peak of the pulses has a hard spectrum, its lower photon index ($\sim -0.5$) exceeds the synchrotron limit of $-2/3$ \cite{1998ApJ...506L..23P}. The hardening of a spectrum can be caused by the Compton up-scattering process, that a partial synchrotron (or jitter) photons act as the seed photons and are upscattered to the GeV energy range. To obtain a photon index as low as $\sim -0.5$, it is essential to change the distribution of the photons with energy above the Klein-Nishina limit, hence the photon energy density shall be far beyond the magnetic energy density \cite{Jiang:201586a,Asano:2009eb8} or/and the high-order Compton up-scatterings becoming important. Such a high photon energy density induces a high opacity of $\gamma\gamma \rightarrow e^+e^-$ collision, the plasma outflow is possibly in a semitransparent state for GeV photons during the entire prompt emission. The opacity decreases during the plasma expansion, eventually it becomes transparent then GeV photons escape, a photon with higher energy has a longer delay of the transparency, as observed in figure \ref{fig:190114C_LIV}.  In addition, a photosphere may seat behind, it cools during the expansion and injects thermal photons to the outflow ahead, thermal photons also act as seed photons, possibly even contribute the majority of the seed photons. They increase the GeV photons by the Compton up-scattering (softening the GeV spectrum), the gamma-ray photons are increased via the cascaded particles (hardening the gamma-ray spectrum). And consequently, a positive correlation of gamma-ray and GeV fluxes, and a negative correlation of the gamma-ray and GeV spectral indices may exhibit, as observed in figure \ref{fig:190114C_LAT_GBM_Evolution}c and figure \ref{fig:190114C_LAT_GBM_Evolution}d, a qualitative demonstration of this procedure is shown in figure \ref{fig:CartoonSpectrumBoosted}. The prominent role that the thermal injection plays is supported by figure \ref{fig:190114C_thermal}, that the observed temperature keeps decreasing while a high percentage of $\sim 30\%$ of thermal preserves. The existence of a semitransparent state, the consequent intense attenuation and cascade of the GeV photons during the prompt emission are supported by figure \ref{fig:190114C_LAT_GBM_Evolution}c, that during the peak of pulses, the ratio of GeV flux to gamma-ray flux\footnote{In figure \ref{fig:190114C_LAT_GBM_Evolution}, the GeV flux ranges from $100$~MeV to $1$~GeV, and the gamma-ray flux ranges from $1$~keV to $10$~MeV} has a value ($\sim 4\%$), which is much lower than the corresponding value ($\sim 25\%$) between the two prompt pulses ($\sim 2.5$~s) and in the phase of post prompt emission ($>30$~s).

Poynting flux is another candidate of the GRB outflow, it is natural to produce a broken-power-law-like spectrum, with a hard low-energy part, and a high-energy part extending to the GeV, all by synchrotron emission\cite{Zhang:2010564}. The GeV delay is proposed to be caused by the optical depth of the photon-photon pair creation \cite{Bosnjak:2012705}. But Poynting flux is not in favour of the high percentage of thermal emission. Relativistic protons are sources of GeV and TeV photons, by injecting power-law distributed protons with an index $-2$ and a cutoff energy at $\sim 10^{21}$~eV, it is possible to produce the GeV and TeV photons forming a power-law-like spectrum distribution with photon index $\sim -1.8$, as we observed at $\sim 8000$~s \cite{2010PhRvD..82d3002A}. But for producing the observed GeV flux in the prompt emission, it requires a very high proton isotropic energy $>10^{55} {\rm erg~s}^{-1}$ \cite{Asano:2009eb8}. And for producing the GeV emission in the afterglow phase by the traditional forward shock acceleration model, it needs enormously high kinetic energy $\sim 10^{58}$~erg \cite{Fan:20132be}. The proposal of accelerating the protons by the strong electromagnetic field encompassing the black hole may help extricate from the energy dilemma, and it is able to generate the prompt and the afterglow GeV emissions by the infalling supernova remnant \cite{2018arXiv181101839R}.




The photon index of GRB 190114C is very soft, it breaks the synchrotron fast-cooling limit ($-3/2$) \cite{1998ApJ...497L..17S} after the pulses, in additional to the break of the synchrotron limit ($-2/3$) during the first pulse. It is rare of observing both breaks in a single GRB, especially they occur in sequences of time bins and are confirmed by data of high signal to noise ratio ($>20$). This result infers the GRB has entered the afterglow phase after the pulses, and the cutoff energy is found to be $<100$~keV after $50$~s. 

The time delay of GeV spectrum puts a limit on the effective quantum-gravity energy scale of Lorentz invariant violation (LIV), for a linear effect, the energy scale is $E_{\rm QG,1} > 0.63\times10^{16} ~ \rm GeV$, and for the quadratic effect, the energy scale is $E_{\rm QG,2} > 3.4\times10^{7} ~ \rm GeV$, these value are consistent with the constraints from the spectral delay of GRB 160625B \cite{2017ApJ...834L..13W}. For individual photons, higher energy photon has a longer delay, as found $T \propto E^{1/3}$ in figure \ref{fig:190114C_LIV}, here $T$ is the delayed time and $E$ is the photon energy. This relation of index $1/3$ infers the GeV delay is not a pure LIV effect. If we attribute this relation to the energy dependency of the transparent time, it infers the outflow is not in a pure radiation dominated scenario, matter dominated scenario or Poynting flux dominated scenario. A realistic outflow may contain many compositions, and the GeV delay is caused by a conjugation of several effects. Besides, by extrapolating the time delay relation in figure \ref{fig:190114C_LIV}, the first $1$~TeV photon may be observed at $\sim 35$~s, therefore, a TeV spike related to the prompt emission coule be appear in the MAGIC observation, then followed by a power-law decay. 

To close this article, we quote two sentences that we truly recognise, the first sentence is written by Johan P. U. Fynbo \cite{Fynbo:2014e82}, it was for the legend GRB 130427A, and it suits the successor GRB 190114C, ``\emph{there are challenges to understanding all the details in this amazing dataset. This is good news because it leaves us with the expectation that Nature has here given us precious new clues on how to improve our models and reach a deeper understanding of the end life of massive stars}''. The second one is cited from the latest GRB book written by Bing Zhang \cite{zhang_2018}, ``\emph{due to their elusive nature and the technological challenges in observing them, GRBs have not been observed in all wavelengths at all epochs. As a result, this field has been and will remain a hot subject in contemporary astrophysics}''.

\noindent

\bibliographystyle{Science.bst}
\bibliography{190114C}

\clearpage

\section*{Acknowledgements}
We acknowledge the use of the public data from Fermi and Swift data archives. We appreciate the discussion with Prof. She-sheng Xue, Prof. Gregory Vereshchagin, Prof. Razmik Mirzoyan and Dr Mile Karlica on the high energy emission. We especially thank Dr Ruoyu Liu with whom fruitful discussions were carried out during many mid-nights.


\section*{Author contributions}
YW initiated the paper, performed \emph{Fermi}-LAT and \emph{Swift}-XRT analysis, compared GRB 130427A and GRB 190114C, found the correlation of gamma-ray and GeV emissions, and proposed the theoretical interpretation. LL was in charge of the \emph{Fermi}-GBM analysis and the thermal component, YW and RM assisted. YW wrote the majority of the article, LL wrote the sections of \textit{Fermi}-GBM and thermal component, RM assisted. YW, LL, RM and RR participated in the discussion.

\section*{Competing interests}
The authors declare no competing interests.

\section*{Data and materials availability}
All data used in this paper are retrieved from the public data of \emph{Fermi} and \emph{Swift} satellites, the fitting results are provided in the supplementary material (Table S1 - S4),

\noindent
\textbf{\Large List of Supplementary materials}\\
\noindent
{\large Materials and Methods S1 - S11\\
Figures S1 - S10\\
Tables S1 - S4\\
References (51 - 80)\\
}\\


\setcounter{figure}{0}    
\setcounter{section}{0}
\setcounter{table}{0}
\renewcommand{\thesection}{S\arabic{section}}
\renewcommand{\thefigure}{S\arabic{figure}}
\renewcommand{\thetable}{S\arabic{table}}
\renewcommand{\theequation}{S\arabic{equation}}

\noindent
\section*{Materials and Methods}

\section{Satellites Time Coverage} 
\label{sec:satellites_time_coverage_and_test_statistics}

\begin{figure}[ht]
\centering
\includegraphics[width=1\hsize,clip]{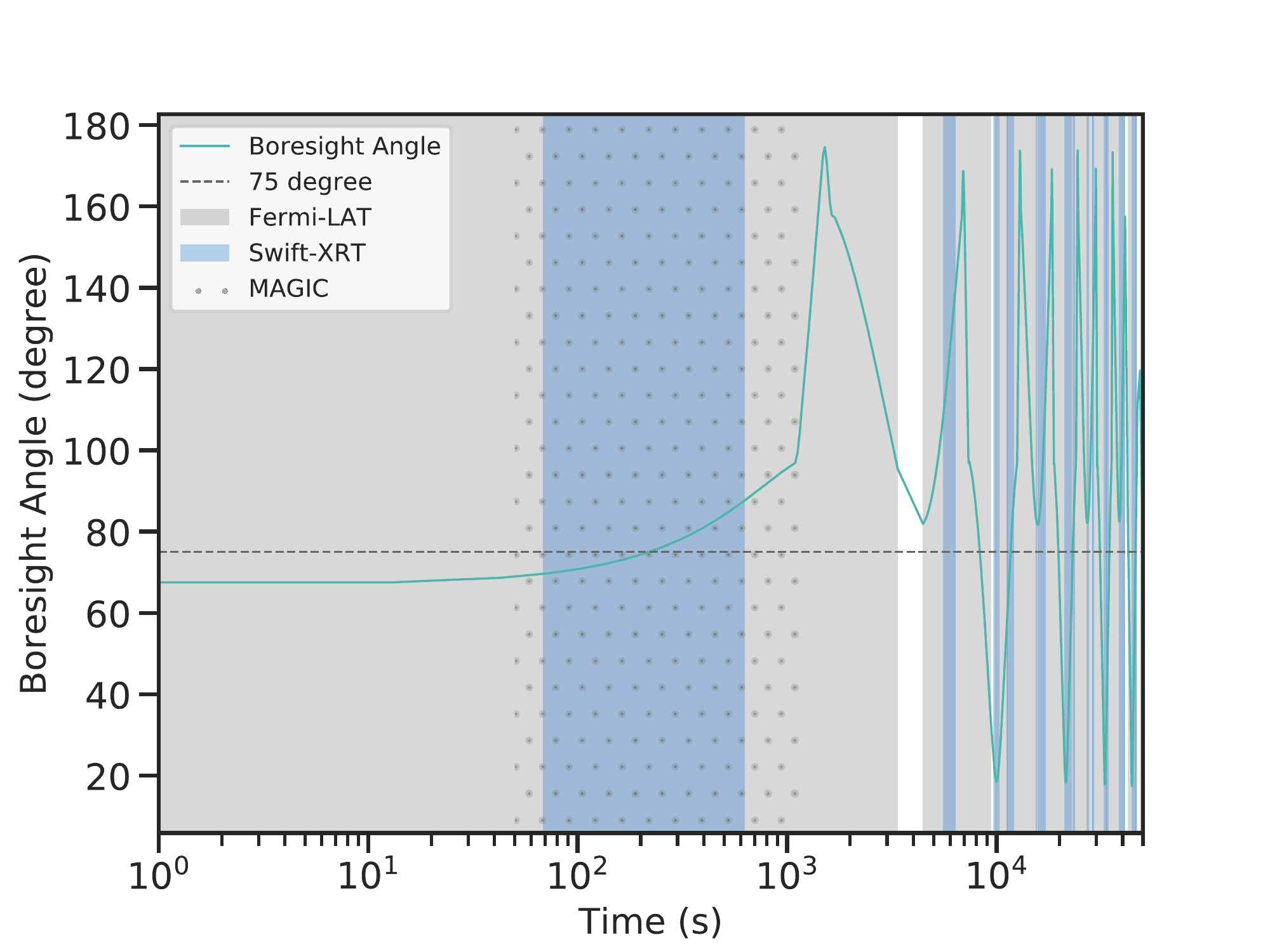}
\caption{The bore-sight angle (the angle between the GRB 190114C and the Fermi-LAT bore-sight) of \textit{Fermi}-LAT, as well as the time coverage of \textit{Fermi}-LAT, \textit{Swift}-XRT and MAGIC. The green curve is the evolution of the bore-sight angle since the trigger, the grey dotted line indicates $75$ degree, outside which the effective area of \textit{Fermi}-LAT becomes very small. The grey shades cover the good time intervals of the \textit{Fermi}-LAT observation. The blue shadow and the grey dots correspond to the observation time of  \textit{Swift}-XRT and MAGIC.  There exists a common time interval from $\sim 50$~s to $\sim 200$~s covered by Fermi and MAGIC, as well as partially covered by Swift and other lower energy telescopes.}
\label{fig:bore-sightAngle}
\end{figure}

The effective area and the point spread function of Fermi-LAT highly depend on the off-axis angle from the Fermi-LAT bore-sight to the position of the GRB,  it indicates if the GRB is well located within the field of view (FOV), hereafter we call it bore-sight angle. The higher bore-sight angle leads to fewer GRB photons observed, a typical threshold of bore-sight angle is $75$ degree, outside of which the probability of determining the photons becomes tiny. GRB 1901114C has an initial bore-sight angle of $67$ degree, shown in figure \ref{fig:bore-sightAngle}. For $10$~GeV photons, a $67$ degree angle corresponds to an effective area of $\sim 0.2 ~\text{m}^2$, which is only $\sim 20\%$ of the effective area when the satellite pointing to the GRB \footnote{\url{http://www.slac.stanford.edu/exp/glast/groups/canda/lat_Performance.htm}}. The bore-sight angle keeps increasing, at time $\sim 200$~s, it passes the threshold of $75$ degree, then re-enters at the time later than $\sim 8000$~s when the GeV emission is supposed to weak. Therefore, we may expect the majority of Fermi-LAT photons are received within $300$~s since the trigger time. The MAGIC observation starts $50$~s after the Fermi-GBM trigger, and lasts $20$ minutes, there exists a common time interval from $\sim 50$~s to $\sim 200$~s covered by Fermi and MAGIC, as well as partially covered by Swift and other lower energy telescopes, the time coverage of Fermi, Swift and MAGIC is shown in figure \ref{fig:bore-sightAngle}.

\section{Test Statistics of Fermi-LAT}
\label{sec:test_statistics_of_fermi_lat}
\begin{figure}[ht]
\centering
\includegraphics[width=1\hsize,clip]{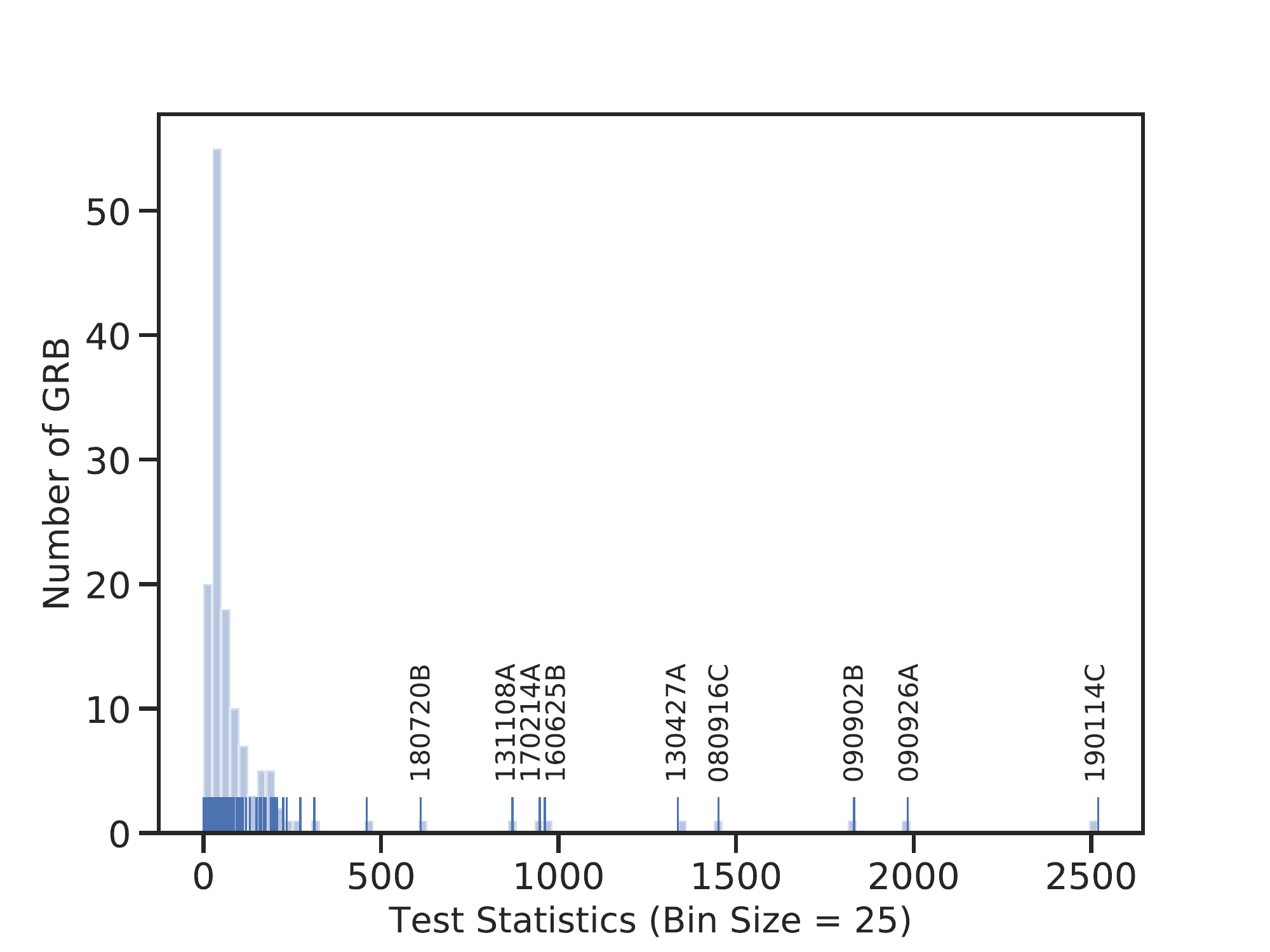}
\caption{Test Statistics (TS) of GRB 190114C and other 138 GRBs observed by Fermi-LAT (\url{https://fermi.gsfc.nasa.gov/ssc/observations/types/grbs/lat_grbs/table.php}). There are 9 GRBs with significant $\text{TS}>500$.  Here for computing the TS value of GRB 190114C, we consider the photon energy form $100$~MeV to $100$~GeV within T90, from $0-116$~s, A TS value of $2520$ (shown in the figure) is obtained. For $0 - 200$~s, the period when the GRB is inside the field of view, the TS value is $2451$. GRB 190114C is one of the most significant GeV GRBs, if not the top one.}
\label{fig:TS_Histogram}
\end{figure}

Despite having a large bore-sight angle, still, GRB 190114C challenges the record of observing the most significant GeV photons. The significance is determined by the value of Test Statistics (TS) computed from the Likelihood Ratio Test (LRT). LRT compares two models, one considers only the background, the other contains an additional putative GRB as a point source, the ratio of the likelihoods by fitting these two models gives a TS value, a higher TS value presents more confidence in having an additional GRB model, $\text{TS} = 25$ corresponds to $\sim 5$ sigma, details see \cite{2013ApJS..209...11A}. Here we perform the Fermi-LAT data likelihood analysis using Fermitools distributed as a Python package \footnote{\url{https://github.com/fermi-lat/Fermitools-conda}}, we choose the response function file \textit{p8\_transient020}, energy range is from $100$~MeV to $100$~GeV, and a region of $12$~degree around the source is taken into the analysis. The TS value we obtained for the T90 from $0 - 116$~s is $2520$, which is extremely high, perhaps the highest among all the Fermi-LAT GRBs, normally the TS value is smaller than $250$ if there is GeV detection, as shown in figure \ref{fig:TS_Histogram}\footnote{The value of TS depends on the selection of time range and energy range. The TS value obtained from the link of this figure considers a common $100$~MeV to $10$~GeV energy range, with different time ranges.}. The TS value for the time of $0-200$~s,  when the GRB is within the FOV of the satellite, is $2451$.

\section{GeV Photons}
\label{sec:gev_photons}

\begin{figure*}[ht]
\centering
\includegraphics[width=1\hsize,clip]{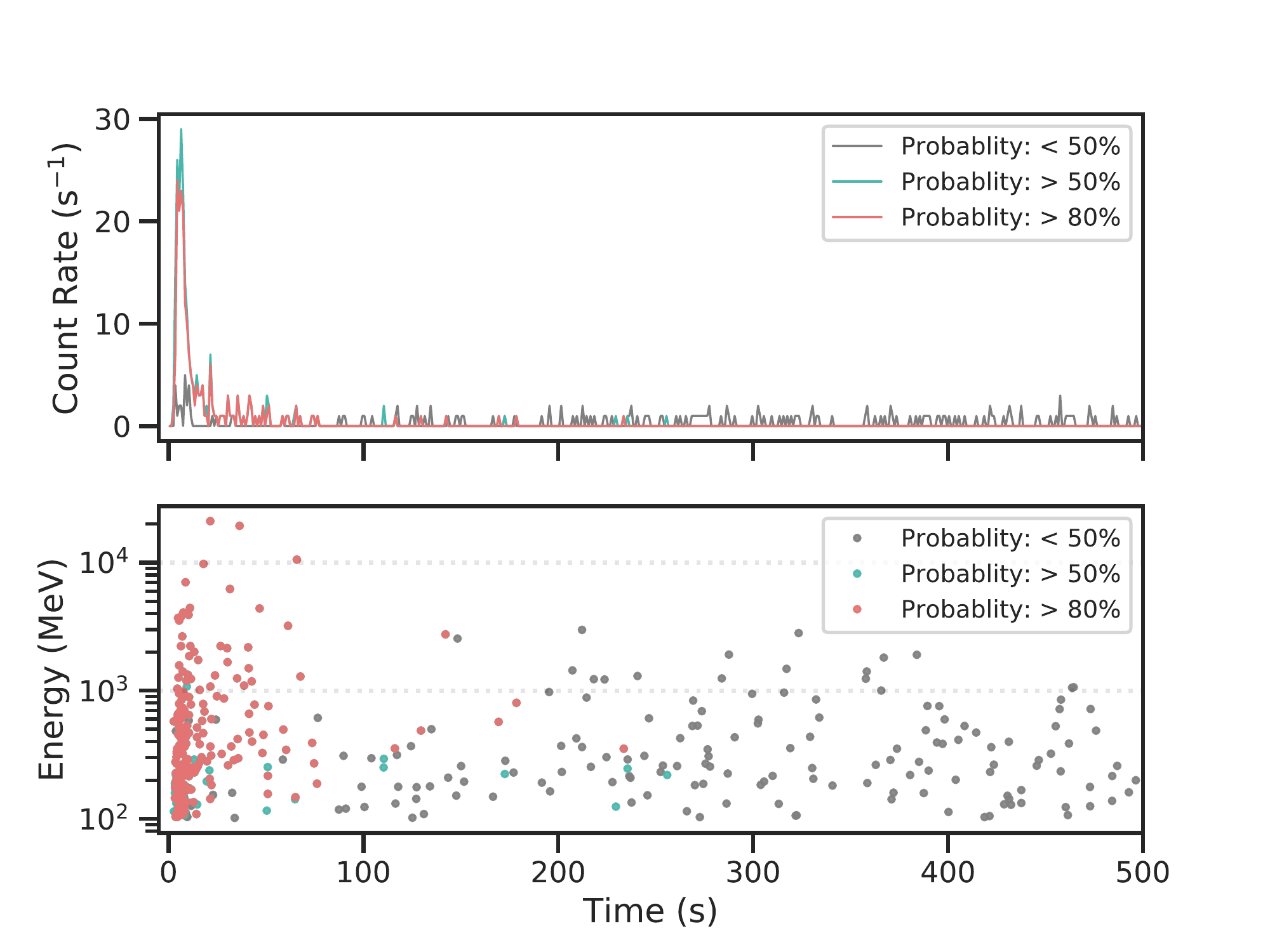}
\caption{The \textit{Fermi}-LAT photons in the first $500$~s. Upper: the count rate light-curve for photons of energy $100$~MeV to $100$~GeV. Lower: the energy of each single photon and its arrival time. Grey, red and blue colours represent $<50\%$, $>50\%$ and $>80\%$ of the probability that this photon belonging to the GRB 190114C. After $250$~s, there is no photon having high probability $>80\%$.}
\label{fig:190114C_GeV_Photons}
\end{figure*}

High TS value signifies a high abundance of GeV photons, in figure \ref{fig:190114C_GeV_Photons}, we show the probability of each photon belonging to this GRB 190114C and the count rate light-curve. There are in total $208$ photons with high probability ($>80\%$), within which, $153$ photons are received before $20$~s, the highest rate reaches more than $20 s^{-1}$. There is only one photon comes after $200$~s, this is consistent with our expectation that the rate of GeV photons decays with time, and the GRB goes outside the $75$~degree of the boresight angle after $200$~s. The first high energy photon ($>100$~MeV) received with probability $>80\%$ is at $7.33$~s, and with probability $>50\%$ is at $2.74$~s. The photon with the highest energy $21.05$~GeV arrive at $21.43$~s, besides there are two more photons with energy more than $10$~GeV, they are of energy $19.36$~GeV and $10.54$~GeV, arriving at $36.47$~s and $65.86$~s respectively.

\section{Fermi-LAT Data Analysis}
\label{sec:fermi_lat_data_analysis}

The data of \emph{Fermi} satellite is obtained from the Fermi Science Support Centre \footnote{\url{https://fermi.gsfc.nasa.gov}}, and analysed by Fermitools \footnote{\url{https://github.com/fermi-lat/Fermitools-conda/wiki}}. Spectrum is fitted by the unbinned likelihood analysis, taking into the consideration of background point-like sources, Galactic diffuse and isotropic emission, we adopt the response function of \emph{P8$\_$TRANSIENT020} for transient sources. The time-integrated Fermi-LAT spectrum of first $200$~s is best fitted by a single power-law, with a typical photon index $-1.98\pm0.062$, the averaged flux is $1.38\pm0.20 \times 10^{-6} ~\text{erg} ~\text{s}^{-1} \text{cm}^{-2}$, corresponds to the isotropic Fermi-LAT energy of $1.31\pm0.19 \times 10^{53}$~erg in the energy band of $100$~MeV to $100$~GeV, k-correction is considered. The time-resolved spectral fitting is performed from the trigger time to $10^5$~s. Time bins are spitted by the Bayesian block method, with an additional criteria that, the TS value of each bin shall be between $100$ and $200$ before $50$~s, and at least $20$ at later time, we give the upper limit of the flux if the TS value of the time bin is smaller $20$. In figure \ref{fig:190114C_index}, we show the evolution of the photon index and the energy flux,  the initial power-law index is very soft as $-3.65\pm0.48$, it rises then varies around $-2$, the photon index shows an almost anti-flux-tracking behaviour after$\sim 4.5$ ~s. At time $\sim 8000$~s,  the GRB re-enters the FOV of Fermi-LAT, the TS value of this time bin is $25$, which makes possible to obtain the spectral fitting, a power-law index of $-1.72\pm0.34$ is obtained. We also tested cut-off power-law and broken power-law to fit the spectra, but none of them makes a significant improvement,  a single power-law is statistically always preferred, even for the bins with very soft index. 

With the spectral parameters from the time-resolved fitting, and the detected redshift ($z=0.42$), we are able to generate the luminosity light-curve of Fermi-LAT.  In figure \ref{fig:190114C_luminosity}, we show two light-curves, presenting the luminosity of $100$~MeV to $1$~GeV and $100$~MeV to $100$~GeV in the cosmological rest-frame (by default, we use observer's frame if we don't specify), respectively. In the first $2.7$~s, GeV emission has a low TS value $11$, which is consistent with figure \ref{fig:190114C_GeV_Photons}, that there is no confident photon arriving before $2.74$~s, an upper limit is computed. Then the GeV light-curve exhibits spiky structure till $25$~s, as like the prompt gamma-ray emission. Later, the GeV light-curve seems to follow a power-law decay, the power-law index relies on the time bin at $\sim 8000$~s, the power-law index is obtained as $-1.12\pm0.12$ and $-1.16\pm0.14$ for the light-curves of $100$~MeV to $1$~GeV and $100$~MeV to $100$~GeV, respectively. For comparison, we show the X-ray light-curve of Swift-XRT in the energy range of $0.3 - 10$~keV. The power-law decay index of X-ray light-curve from the starting time of XRT observation ($\sim 60$~s) to $\sim 8000$~s is $-1.36\pm0.0015$, which is consistent with other GRBs \cite{2015ApJ...805...13L,2018ApJS..236...26L}, and slightly steeper than the Fermi-LAT light-curve.

\section{Fermi-GBM Data Analysis}
\label{sec:fermi_gbm_data_analysis}

\begin{figure}[ht]
\centering
\includegraphics[width=1\hsize,clip]{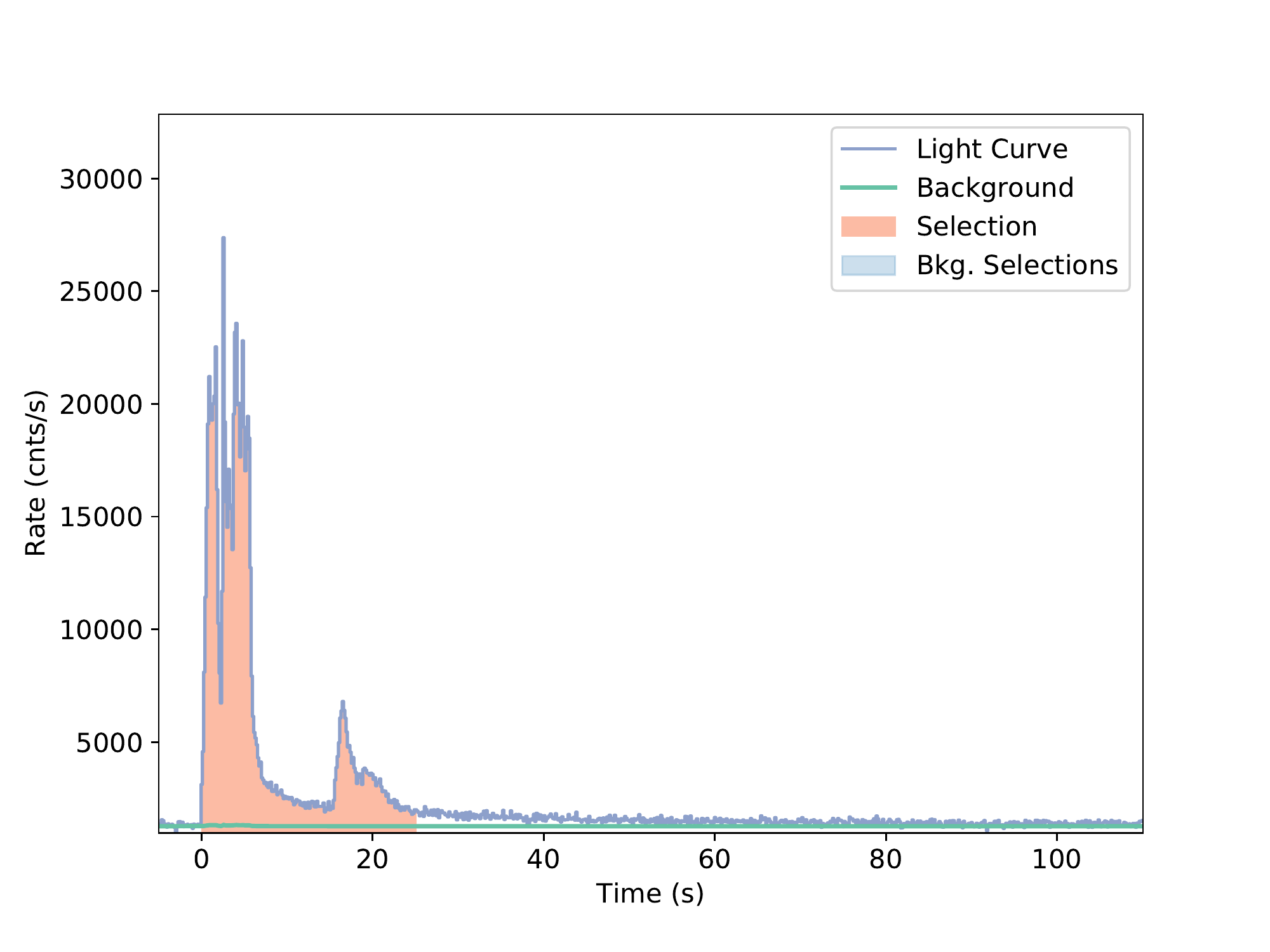}
\caption{Count light-curve of \textit{Fermi}-GBM. This plot covers the entire T90 of $0-116$~s. The background is fitted by the green line. The orange shadow denotes the two pulses.}
\label{fig:190114873_lc}
\end{figure}

The temporal and spectral analysis for \textit{Fermi}-GBM data is carried out by using the fully Bayesian approach analysis package, namely, the Multi-Mission Maximum Likelihood Framework (3ML) \cite{2015arXiv150708343V}). The GBM carries 12 sodium iodide (NaI, 8keV-1MeV) and 2 bismuth germinate (BG0, 200 keV-40 Mev) scintillation detectors \cite{2009ApJ...702..791M}.
Four brightest NaI detectors (n3, n4, n7, n8) and one BGO detector (B0) are used to conduct the analysis. 
The background is fitted with automatically determined polynomial order (0 and 1, marked with 'Background' in Figure \ref{fig:190114873_lc}) by the 3ML, before (-20 s to -10 s) and after (180 s - 200 s) the burst (marked with 'Bkg.Selections'). We select the source as the time interval from -1.0 s to 150 s (marked with 'Selection'). The maximum likelihood-based statistics are used, the so-called Pgstat, given by a Poisson (observation)-gaussian (background) profile likelihood \cite{1979ApJ...228..939C}. 
Light curve of prompt emission, with the source selection, the background selections, and the best background fitting,
is shown in Figure \ref{fig:190114873_lc}. The light curve in GRB 190114C consists two clear pulses, with first the main one (up to $\sim$ 15 s) and later the weaker one (from $\sim$ 15 s to $\sim$ 25 s). 

The analysis of time-integrated and time-resolved spectra is the main approach to study the GRB radiation mechanism, which has widely studied in previous works \cite{2006ApJS..166..298K,2018arXiv181003129L}. Considering that the burst is extremely bright, it is interesting to find the model best representing the data. We apply 6 models to fit the time-integrated spectrum for the burst (see Table \ref{tab:promptModel}), respectively.  These models are power-law (PL), cutoff power-law (CPL), Band, PL+BB, CPL+BB, and Band+BB. The model definitions are in supplement section \ref{sec:fitting_functions}. We generally find the CPL+BB (or the Band+BB) fitting is much better than the others, and the Bayesian information criterion is at least 300 less than other models. This is quite surprising and it indicates that adding a thermal component largely improves the spectral fitting. 

We further conduct the time-resolved spectral analysis from $-1.0$~s to $150$~s covering the T90 of $106$~s. We bin the Time-Tagged Events (TTE) data of the brightest NaI detector using the Bayesian Blocks method (BBs) \cite{2013ApJ...764..167S} with false alarm probability $p_{0}$=0.01, and other detectors follow the same bin times \cite{2018arXiv181003129L}. We obtain $49$ spectra in total. We first use the typical GRB spectral model the CPL model to fit the data, to obtain the temporal evolution of the low energy index $\alpha$. We find the $\alpha$ evolution presents soft-to-hard trend in the first $4$ seconds, even across the synchrotron limit ($-2/3$) \cite{1998ApJ...506L..23P}, then displays a hard-to-soft evolution, at late time it is softer than the fast-cooling limit ($-3/2$) of the synchrotron model \cite{1998ApJ...497L..17S}, which is quite surprising.

In view of the very hard spectral index in the time-resolved spectrum, we are motivated to add the BB component to fit the time-resolved spectrum. Indeed, we find adding a BB component fits better for those time-resolved spectra which display hard $\alpha$ spectral indices. The BB component is confidently detected around $\sim 1.5$~s, and from $\sim$ 2.5 s to $\sim$ 5.7 s, covering the first pulse. The temperature evolution is shown in figure \ref{fig:190114C_thermal}, and an example fit data in one time bin (between $2.57$ s to $3.63$ s) of GRB 190114C using the CPL plus the BB component is presented in figure \ref{190114C_spectrum_gbm_257_363_lat_583_688_XSPEC}.

\section{Swift-XRT analysis}
\label{sec:swift_xrt_analysis}
The data of Swift-XRT are retrieved from UKSSDC \footnote{\noindent \url{http://www.swift.ac.uk}} and are analyzed by NASA's Heasoft 6.25 with relevant calibration files\footnote{\noindent \url{http://heasarc.gsfc.nasa.gov/lheasoft/}}. By applying the criteria of binning the data, that each time bin contains at least $1000$ photons, as well as the signal to noise ratio must be larger than $3$. We obtain $43$ time bins in total, including $20$ from the windows timing mode, and $23$ bins from the photon counting mode, they cover the time, in the observer's frame, from $68.13$~s to $303992.39$~s since the Fermi-GBM trigger. Following the standard procedure of Swift-XRT data analysis \cite{Evans:2007iz,Evans:2009kx}, we fit the spectra from $0.3$~keV to $10$~keV of all the time bins by a single power-law model considering the X-ray absorption from our Galaxy and the progenitor's galaxy. To produce the luminosity light-curve, we applied the k-correction to each time bin, the k-correction converts the flux observed on the earth that reshifted by the expansion of the universe to the progenitor's cosmological rest frame, k-correction is computed by the redshift and the model's fitted parameters \cite{2001AJ....121.2879B,2018ApJ...852...53R}.

\section{Fitting Functions}
\label{sec:fitting_functions}

The Band function \cite{1993ApJ...413..281B} is defined as
\begin{eqnarray}
\label{n}  N_{E}=A \left\{ \begin{array}{ll}
(\frac{E}{E_{\rm piv}})^{\alpha} e^{-E/E_{0}}, & (\alpha-\beta)E_{0}\geq E \\
(\frac{(\alpha-\beta)E_{0}}{E_{\rm piv}})^{(\alpha-\beta)} e^{(\beta-\alpha)}(\frac{E}{E_{\rm piv}})^{\beta}, & (\alpha-\beta)E_{0}\leq E \\
\end{array} \right.
\end{eqnarray}
where $N_{E}$ is the photon flux ($\rm ph/cm^{2}/keV/s$), $A$ is the normalisation for the spectral fit, $E_{\rm piv}$ is the pivot energy fixed at 100 keV, $E$ is the photon energy, and $E_{0}$ is the break energy. The function consists of two power laws which are smoothly separated by a peak energy $E_{\rm pk}$ (in the $\nu F_{\nu}$ space in units of keV), described by the low-energy photon-index $\alpha$, and the high-energy photon index $\beta$.

The Comptonized model (COMP) is the power-law model with a high-energy exponential cutoff:
\begin{equation}
N_{E} = (\frac{E}{E_{\rm piv}})^{\alpha} e^{-E/E_{0}},
\label{Eq:CPL}
\end{equation}

The Planck function (BB) is given by the photon flux
\begin{equation}
N_{E}(E,t)=A(t)\frac{E^{2}}{exp[E/kT(t)]-1},
\end{equation}
where $E$ is the photon energy, $k$ is the Boltzmann constant. 
It has two free parameters: $T=T(t)$ and the normalisation $A(t)$ of the photon flux.

A power-law function (PL), 
\begin{equation}
N_{E} =A E^{s},
\label{PL}
\end{equation}
where $A$  is the normalization of the photon flux and $s$ is the power-law index.

\section{Example of Spectrum Fitting by Markov Chain Monte Carlo}
\label{sec:example_spectrum_fitting_t90}
\begin{figure*}
    \centering
    \includegraphics[width=1\hsize,clip]{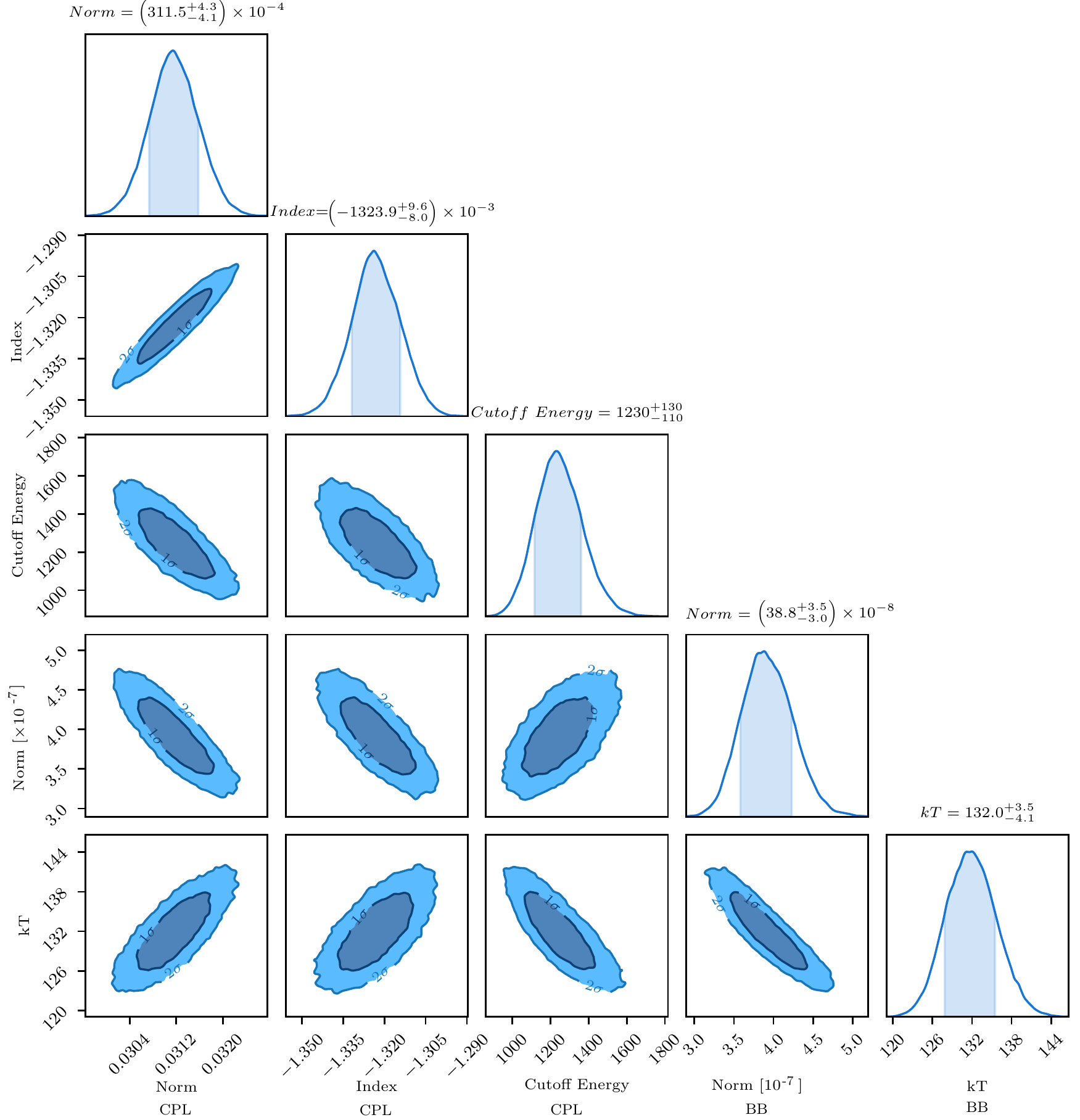}
    \caption{Bayesian Monte Carlo fitting of \textit{Fermi}-GBM spectrum from $0$~s to $116$~s (T90). We apply $20$ chains, each chain iterates $10^4$ times and burns the first $10^3$ times. The parameters are normalisation (Norm CPL), cut-off energy and power-law index of the cut-off power-law model, as well as normalisation (Norm BB) and temperature ($k$T) of the blackbody model.}
    \label{fig:datafitting}
\end{figure*}

\begin{figure*}
    \centering
    \includegraphics[width=1\hsize,clip]{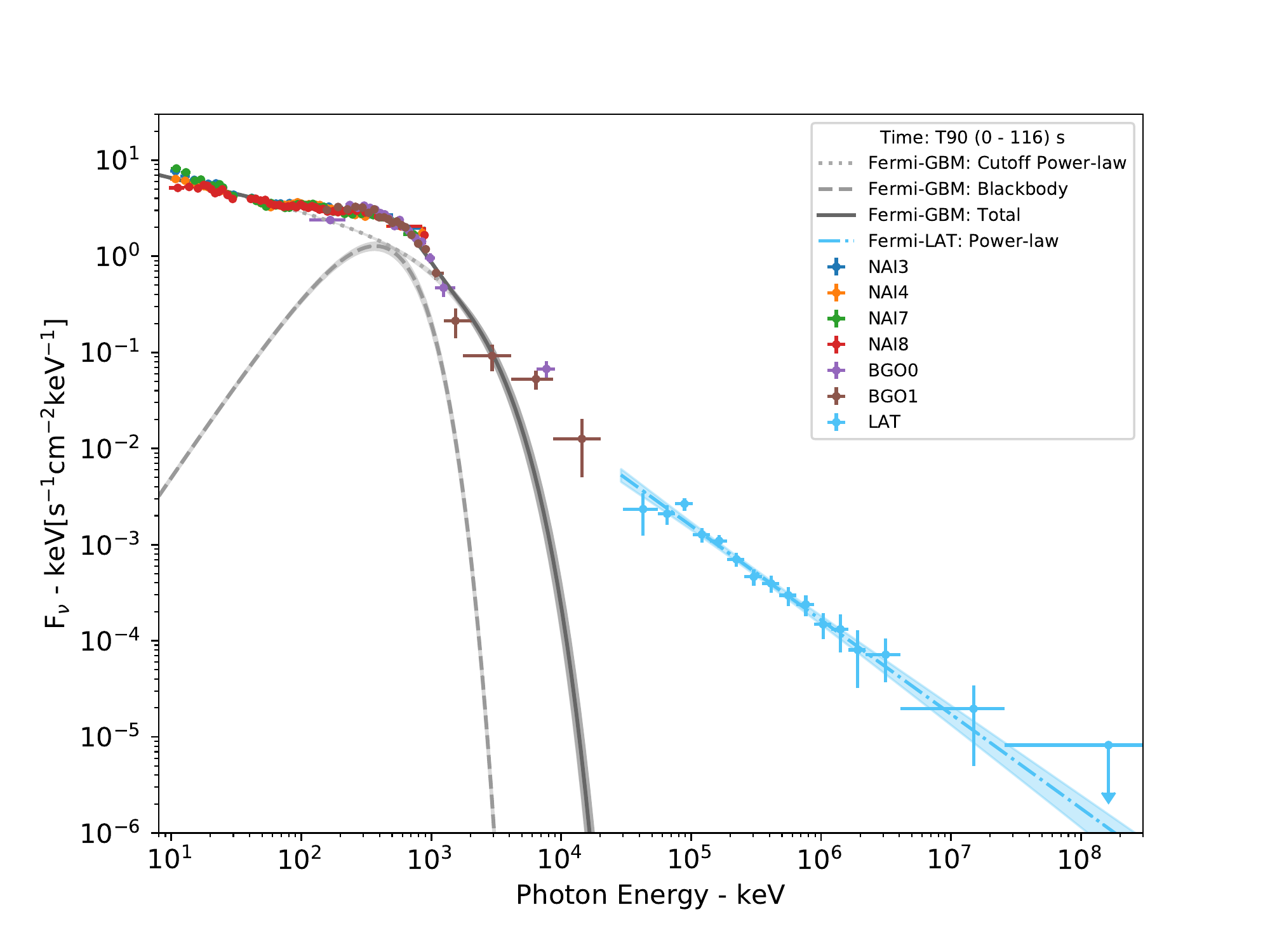}
    \caption{Multi-wavelength spectrum from $0$ to $116$~s (T90). The spectrum includes data from \emph{Fermi}-GBM (4 NAI and 2 BGO detectors) and \emph{Fermi}-LAT. The fitting of Fermi-GBM is presented by a solid line, including the components of a blackbody function in dashed line and a cutoff power-law in dotted line. The \textit{Fermi}-LAT data as blue crosses are fitted by a power-law function as dash-dot line. The extrapolation of \textit{Fermi}-LAT fitting connects the Fermi-GBM data. The spectrum after shifting $3.25$~s is not shown, since it changes little on the time-integrated spectrum of a much wider duration of $116$~s.}
    \label{fig:190114C_spectrum_gbm_t90}
\end{figure*}

Spectra of \emph{Fermi}-GBM are fitted by Bayesian iterations of Markov Chain Monte Carlo using a Python package \emph{Multi-Mission Maximum Likelihood framework}  (3ML) \cite{2015arXiv150708343V}. The AIC is preferred for comparing non-nested models, and BIC is preferred for nested models \cite{10.2307/2291091}. Log(likelihood) is adopted by the method of maximum likelihood ratio test which is treated as a reference of the model comparison \cite{10.2307/1912557}. Here we give an example of fitting the entire $T90$ data of \emph{Fermi}-GBM, that we first fit the spectrum by several different models, fitting parameters are shown in table \ref{tab:promptModel}, which indicates the best model is cutoff power-law plus blackbody, the corresponding corner plot is shown in figure \ref{fig:datafitting}, and the multi-wavelength spectrum is shown in figure \ref{fig:190114C_spectrum_gbm_t90}.

\section{Integrate GeV Energy by Locally Weighted Regression} 
\label{sec:fermi_lat_energy}

\begin{figure}[hb]
    \centering
    \includegraphics[width=1\hsize,clip]{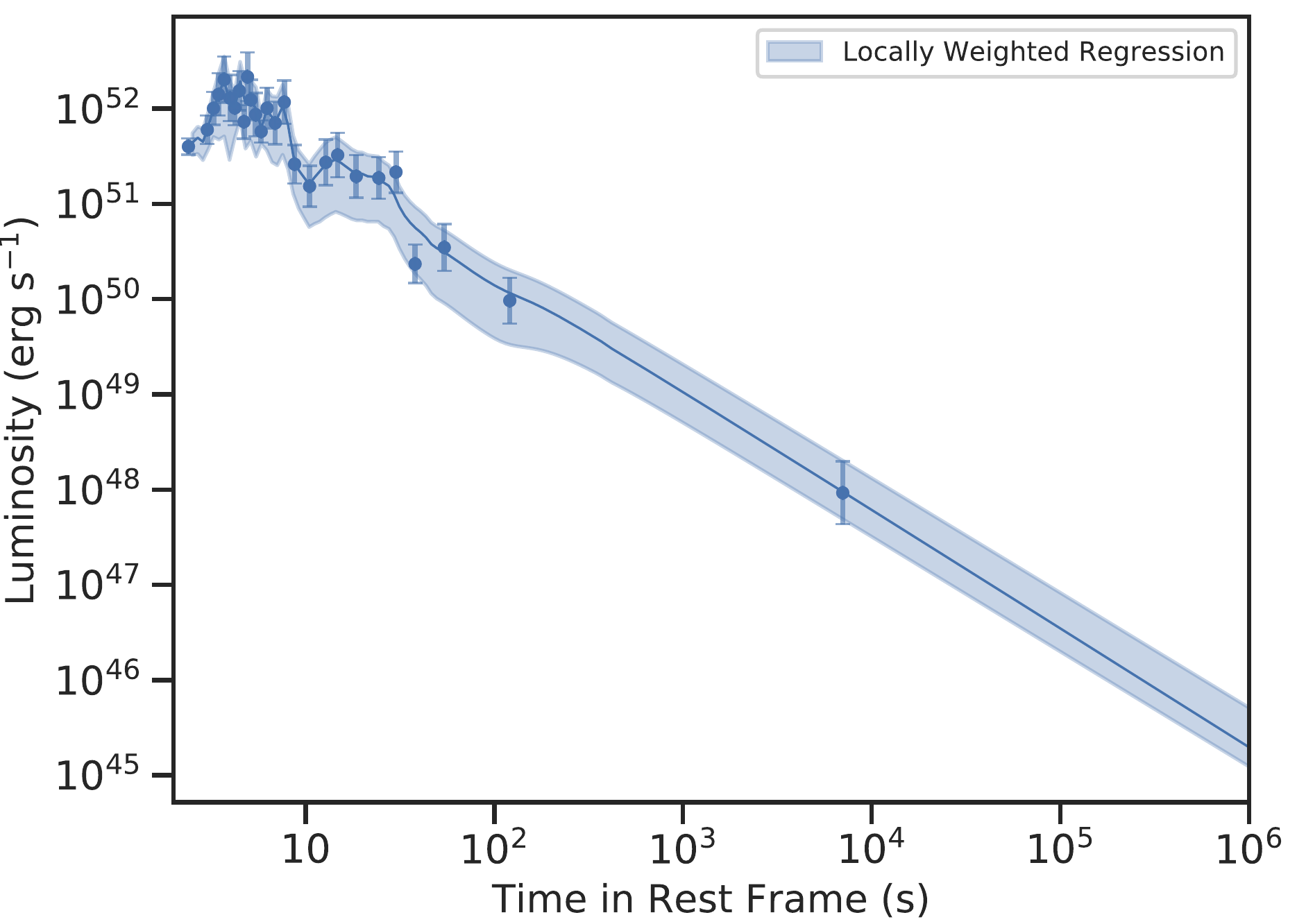}
    \caption{Fitting of Fermi-LAT light-curve by the locally weighted regression. The total energy of \textit{Fermi}-LAT is obtained by integrating the solid line, the shadow is the 1-sigma uncertainty region.}
    \label{fig:LWR}
\end{figure}

The \emph{Fermi}-LAT emission lasts for a long time, presenting in both the prompt emission and the afterglow. The time-coverage of \emph{Fermi}-LAT data is discrete and, especially scarce in the afterglow. To obtain the total energy, we need to integrate the luminosity light-curve, including those time intervals when the data is missing. Here we adopt an algorithm from machine learning named locally weighted regression (LWR) \cite{10.2307/2289282}. LWR predicts the luminosity at a given time by fitting the observed data points with different weights, which depends on the time difference. Therefore, we eventually obtain a sequence of segmental power-law functions constructing the entire light-curve. Summing up all of the segmental power-laws gives the total energy. By applying this method, the energy of 190114C in the energy band $100$~MeV to $100$~GeV during the time from the trigger to $8000$~s is $1.8^{+1.3}_{-1.0} \times 10^{53}$~erg, and the fitting is shown in figure \ref{fig:LWR}.

\section{Cross-Correlation of Gamma-ray and GeV Emission}
\label{sec:cross_correlation}

\begin{figure}[ht]
\centering
\includegraphics[width=1\hsize,clip]{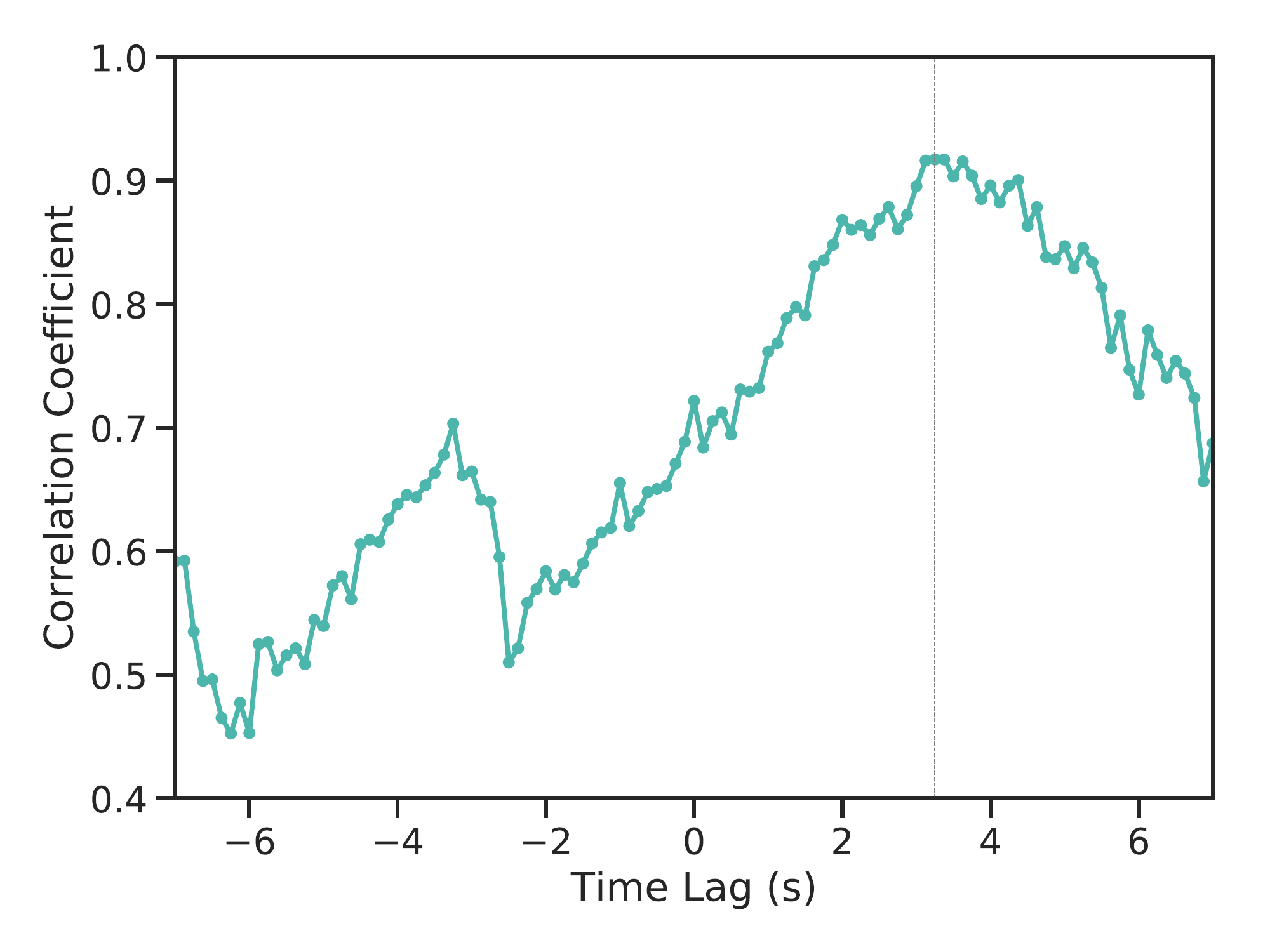}
\caption{Cross-correlation for the flux evolution of gamma-ray and GeV emissions. The x-axis is the time lag of \emph{Fermi}-GBM with respect to the \emph{Fermi}-LAT, the y-axis is the corresponding correlation coefficient. At time lag $3.25$~s, the correlation coefficient reaches its peak of $0.917$, indicating strong correlation.}
\label{fig:190114C_Spectra_DCF}
\end{figure}

\begin{figure}[ht]
\centering
\includegraphics[width=1\hsize,clip]{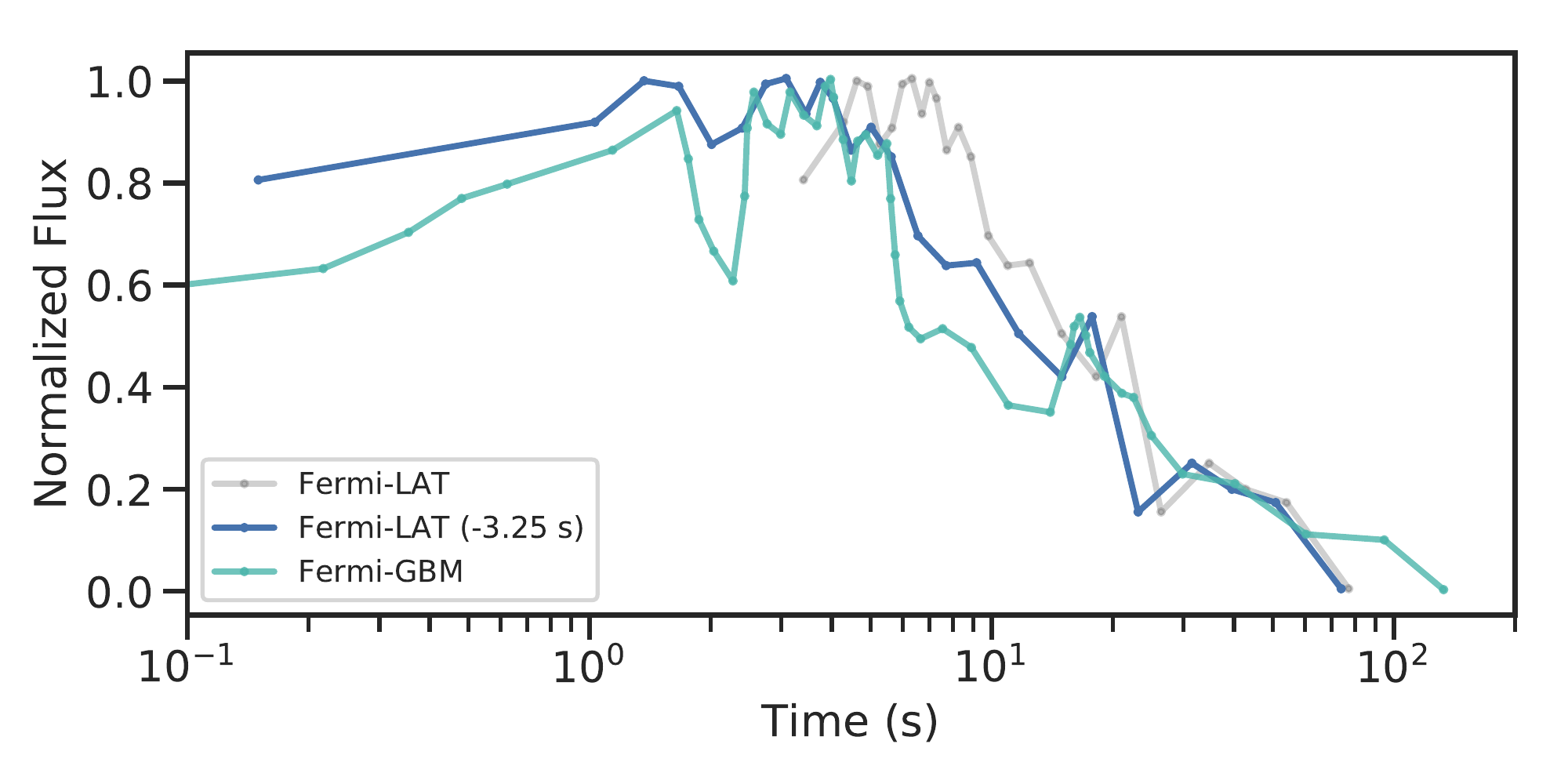}
\caption{Normalised \textit{Fermi}-GBM and \textit{Fermi}-LAT fluxes. The flux is normalised by equation \ref{eq:norm_flux}. Green line is the gamma-ray flux evolution from \emph{Fermi}-GBM, and grey lines is the GeV flux evolution from \emph{Fermi}-LAT. The blue line indicates the GeV flux evolution after shifting $3.25$~s as we obtained from the cross correlation method. Clearly, the shifted GeV flux evolution matches the gamma-ray flux evolution in details. For example, the spikes at $\sim 1.5$~s, $\sim 5$~s and $\sim 15$~s all have counterparts, the general rise and decay of the curve follows each other.}
\label{fig:190114C_flux_norm_LAT_GBM}
\end{figure}

The cross-correlation is a method from signal processing. It is capable to measure the correlation coefficient of two time series as a function of the time displacement of one series to the other. The resulting value of the correlation coefficient is between $-1$ and $1$, correspond to positive correlation and negative correlation, an absolute value $>0.6$ indicates a strong correlation. In this article, we search for the correlation between the temporal flux evolution of gamma-ray and GeV emissions, observed by \emph{Fermi}-GBM and \emph{Fermi}-LAT. The GeV flux is limited to $100$~MeV to $1$~GeV, because the majority of \emph{Fermi}-LAT photons are less than $1$~GeV. This limited energy band gives better fitting for having the flux, and makes the cross-correlation more robust. First, we normalise the flux
\begin{equation}
    f_{Norm} = \frac{Log(f)-min[Log(f)]+0.01}{max[Log(f)]-min[Log(f)]}
\label{eq:norm_flux}
\end{equation}
where $f$ is the original flux, $min$ obtains the minimum value of flux, while $max$ gives the maximum value of flux, $0.01$ is the offset to avoid zero value. Figure \ref{fig:190114C_Spectra_DCF} shows the correlation coefficient as a function of the time-displacement between \emph{Fermi}-GBM and \emph{Fermi}-LAT,  a spike peaks at $3.25$~s with the correlation coefficient $0.917$. It indicates the flux of gamma-ray and GeV emissions are correlated, which can be confirmed visually in figure \ref{fig:190114C_flux_norm_LAT_GBM}, which shows the normalised flux evolution. From the observation, the displacement of $3.25$~s is consistent with the delayed time of GeV emission relative to the gamma-ray emission, the first photon with energy $>100$~MeV arrives at $2.70$~s. The difference of $0.55$~s may due to the missed initial observation, as we discussed in section \ref{sec:twin_grbs}. We also tested the correlation of the gamma-ray and GeV spectral evolution, for which after shifting $3.25$~s, the correlation coefficient reaches $-0.631$, indicating a strong anti-correlation. Moreover, we notice, for the gamma-ray emission the lower photon index evolves tracking the flux. And for the GeV emission, the photon index evolves anti-tracking the flux. These results of trackings are consistent with the above correlations of flux and spectrum.

To be more precise, the anti-correlation of the spectral index is most obvious during the first pulse. The first second (GeV starting time as 0 second) of GeV spectrum is an exception, the GeV spectral has a very soft photon index and it does not anti-correlate with the gamma-ray photon index, as shown in figure \ref{fig:190114C_LAT_GBM_Evolution}. This exception is phenomenologically explained by the time delay of high energy photons, that higher energy photons have longer delay of arriving time as observed in figure \ref{fig:190114C_LIV}. In the initial first second, there are few photons observed in the high-energy tail of \textit{Fermi}-LAT, inducing a steep decay (soft) of spectrum. As the time goes on, the photons of energy $> 1$~GeV starts to popularise, bringing an initial soft to hard transition of the Fermi-LAT spectral index. The influence of such time delays becomes less significant as the time passes.

\section{GeV Delay}

From the correlation of \emph{Fermi}-GBM and \emph{Fermi}-LAT emissions, a GeV spectrum delay of $3.25$~s is confirmed. From figure \ref{fig:190114C_LIV}, the arriving time of all the GeV photons, a trend that higher energy photon arrives later is observed.  Extending the solid line of figure \ref{fig:190114C_LIV}, the first $1$~TeV photon may be observed at $\sim 35$~s, therefore, the MAGIC telescope may observe a TeV spike related to the prompt emission of the GRB, then followed by a power-law decay. 

Following we attempt to interpret the GeV delay by two general reasons, first is from the viewpoint of Lorentz invariance violation, second is from the difference of transparency time. 

\subsection{Lorentz Invariance Violation}
\label{sec:lorentz_invariace_violation}

In some theories of quantum gravity, the energy of a photon has a dispersion relation due to the Lorentz invariance violation (LIV). A framework was initially proposed by \cite{1998Natur.393..763A}, in which a Taylor series expansion presents the energy, the leading term reads

\begin{equation}
E^{2}\simeq p^{2}c^{2}\left[1-s_{\pm}\left(\frac{pc}{E_{{\rm QG},n}}\right)^{n}\right]
\label{eq:dispersion}
\end{equation}
where p is the photon momentum, c is the constant speed of light, and $E_{\rm QG}$ is an effective quantum-gravity energy scale, $n$ indicates the order of expansion. $s = \pm 1$ signifies the positive or negative affection from the change of photon momentum. Therefore, the propagation speed of a photon depends on its energy
\begin{equation}
v=\frac{\partial E}{\partial p}\approx c\left[1-s_{\pm}\frac{n+1}{2}\left(\frac{E}{E_{{\rm QG},n}}\right)^{n}\right]
\end{equation}
Two photon with higher energy $E_h$ and lower energy $E_l$ have different velocities, here $E_h$ and $E_l$ are defined in the observer's frame. If they are generated at a same moment at redshift $z$, taking into account the cosmological expansion, they will arrive at us with a time delay \cite{2008JCAP...01..031J}

\begin{equation}
\Delta t_{\rm LIV}=-\frac{1+n}{2H_{0}}\frac{E^{n}_{h}-E_{l}^{n}}{E_{{\rm QG}, n}^{n}}
\int_{0}^{z}\frac{(1+z')^{n}dz'}{\sqrt{\Omega_{\rm m}(1+z')^{3}+\Omega_{\Lambda}}}
\label{eq:tLIV}
\end{equation}
where the Hubble constant $H_{0}=67.8$ km $\rm s^{-1}$ $\rm Mpc^{-1}$, the cosmological constant $\Omega_{\rm m}= 0.308$, and $\Omega_{\Lambda}=0.692$, value are from Planck mission. 

In GRB 190114C, after aligning the starting time of gamma-ray and GeV emissions, the correlations of gamma-ray and GeV light-curves emerges, as well as the correlation of their spectral evolution. Assuming the correlation indicates the gamma-ray and GeV are generated by the same origin simultaneously, resulting in a possibility that the Lorentz invariance violation causes the time lag between the gamma-ray and GeV emissions. Two spectra have a time lag of $3.25$~s, the GeV spectrum decays as a power-law, and mostly the photons are around $\sim 100$~MeV. It is consistent with the first Fermi-LAT photon observed at $2.70$~s with respect to the Fermi-GBM trigger. Here we adopt the spectral time lag of $3.25$~s to constrain $E_{{\rm QG}, n}$, the gamma-ray spectra has an energy cut at $\sim 500$~keV, it is safe to neglect $E_{l}^{n}$ term in equation \ref{eq:tLIV}. We obtain, considering the lowest $100$~MeV for the GeV spectrum, and the redshift $z=0.42$, for the linear correlation (n=1)

\begin{equation}
E_{\rm QG,1} > 0.63\times10^{16} ~ \rm GeV
\label{eq:linearLIV}
\end{equation}
and for the quadratic correlation (n=2)
\begin{equation}
E_{\rm QG,2} > 3.4\times10^{7} ~ \rm GeV
\end{equation}
This result is consistent with the constraints from the spectral delay of GRB 160625B \cite{2017ApJ...834L..13W}, but orders of magnitudes lower than the lower limit constrained by GRB 080916C \cite{2009Sci...323.1688A} and 090510 \cite{2009Natur.462..331A}. 

We try to infer the constraint by the arriving time of single photons. We take high energy photons associated to GRB 190114C with probability more than $50\%$ before $300$~s, and adopt the widest energy range of \emph{Fermi}-LAT as $30$~MeV to $300$~GeV. All the photons are plotted in figure \ref{fig:190114C_LIV}, a general feature appears that the higher energy photons come later. The linear correlation using $E_{\rm QG,1} = 0.63\times10^{16}$~GeV from equation \ref{eq:linearLIV} presents the dashed line of figure \ref{fig:190114C_LIV}, which locates in the middle of the data points, this may due to $E_{\rm QG,1} = 0.63\times10^{16}$~GeV is inferred from the spectra, which includes all the photons at a given time interval. If we assume the photons on the lower boundary are all originated in the initial first second, and ignore the hypothetical intrinsic delay caused by some acceleration and radiation mechanisms, we relate the difference of the arriving time to the LIV, then we obtain the a phenomenological formula for the LIV by the fitted function of the lower boundary. In figure \ref{fig:190114C_LIV}, the solid line presents such a fitted function, we obtain 
\begin{equation}
    T = 0.35 ({ E/\rm MeV})^{1/3}+0.5~s
\end{equation}
where $0.5$~s may due to the missed initial observation discussed in section \ref{sec:twin_grbs}, T is the arriving time of photon, and E is the energy of photon, here T and E correspond to $t_{\rm LIV}$ and $E_{h}$ in equation \ref{eq:tLIV} following our assumption. The index of $1/3$ differs from the usual dispersion law of equation \ref{eq:dispersion} described by a Taylor expansion. 

\subsection{Time of Transparency}
\label{sec:intrinsic_delay}

The photon-photon collision has several channels, for instance, the dominating $\gamma+\gamma \rightarrow e^{+}e^{-}$, which requires a threshold energy $E_{th}$ for annihilating with the photon of energy $E$,
\begin{equation}
    E_{th}(E) = \frac{(\Gamma m_e c^2)^2}{(1+z)^2 E} = 625 {\rm MeV} (\frac{1}{1+z})^2 (\frac{\Gamma}{500})^2 (\frac{E}{100 {\rm MeV}})^{-1}
\end{equation}
where $z$ is the redshift, $\Gamma$ is the bulk Lorentz factor, energy is defined in the observer's frame. For the interested energy range of $E$ between $100$~MeV and $10$~GeV, the resultant threshold energy $E_{th} \gtrapprox 10$~MeV, which, from the observed spectra of GRB 190114C (see e.g. figure \ref{fig:190114C_spectrum_gbm_t90}), follows a power-law spectrum with photon index $\sim -2$.

The time of a photon with energy $E$ escapes the outflow is determined by its time of transparency $\Delta T(E)$, which is the related to the radius of transparency $R_T(E)$, assuming this radius is much larger than initial radius and the photo-spherical radius, and the bulk Lorentz $\Gamma \propto R^{a}$, we have

\begin{equation}
    \Delta T(E) = \int^{R_T(E)}_0 \frac{1}{2c \Gamma^2} dR  \propto R_T(E)^{1-2a}
\end{equation}
$R_T(E)$ can be obtained by equating the opacity to unit,
\begin{equation}
    \Gamma^{-2} \int_{E_{th}(E)}^{\infty} \sigma(E) n(E, R_T) R_T(E) dE=1
\end{equation}
Considering a power-law spectrum, the count flux density $n(E, R) \propto R^{-2} E^b$, we obtain
\begin{equation}
    R_T(E) \propto E^{\frac{b+1}{2 a b-1}}
\end{equation}
the transparency time finally is given as a function of photon energy\footnote{A detailed derivation is performed in \cite{Bosnjak:2012705}, the same energy dependence of the time delay is obtained.},
\begin{equation}
    \Delta T(E) \propto E^{\frac{b-2a}{2 a b-1}-1}
\end{equation}
we attribute this transparency time as the time delay of GeV photon with energy $E$.

Taking the averaged photon index $b=-2.09$ from the \emph{Fermi}-LAT observation of $3$~s to $10$~s, and the time delay relation $\Delta T(E) \propto E^{1/3}$ found in figure \ref{fig:190114C_LIV}, the index $a = 0.21$ is derived. From the theory, different compositions of the relativistic outflow lead to diverse evolutions of the Lorentz factor $\Gamma$, in the radiation dominated scenario, $\Gamma \propto R$ ($a = 1$), in the matter dominated scenario, $\Gamma$ is a constant ($a = 0$) \cite{Piran:1993790}, and in the Poynting flux dominated scenario, $\Gamma \propto R^{1/3}$ ($a = 1/3$)  \cite{Drenkhahn:2002279,Granot:2011142}. 

None of the three compositions of outflow conforms to the observation, as well as the LIV effect. A realistic scenario may occur in the transitional phase of many compositions, or it conjugates several effects.

There are other model-based proposals related to the GeV delay in the prompt emission, for instance, impact of a relativistic wind on external matter produces GeV photons\cite{Meszaros:19948f1}; the up-scattering of the emission from an expanding cocoon generates GeV emission \cite{Toma:20090f0}; a very high Lorentz factor fireball has dominating MeV emission in the initial seconds \cite{Ioka:2010f24}.

\begin{figure}[ht]
\centering
\includegraphics[width=1\hsize,clip]{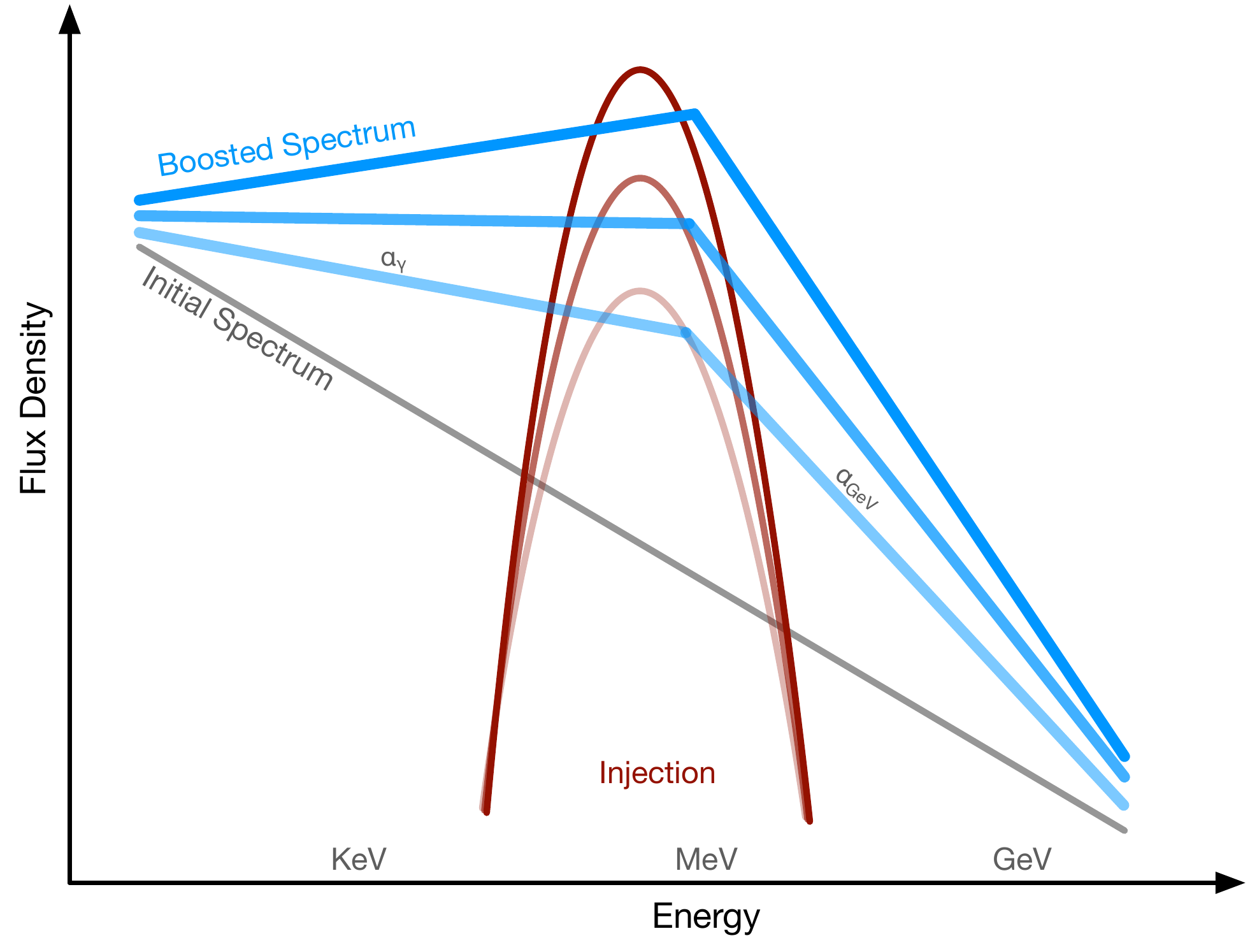}
\caption{A qualitative demonstration of the MeV and GeV spectra, and the anti-correlation of GBM and LAT spectral index. The initial spectrum is presented by a grey line. And an injection (for example, a thermal injection as the red curve) peaking near MeV boosts the initial spectrum, a part of the injected photons are smeared to lower and higher energies by the Compton up-scattering and the consequent cascade. This procedure hardens the KeV-MeV spectrum and softens the GeV spectrum. It is also possible that the thermal seed photons have a higher total number of particles than the electrons, the resultant spectrum shape therefore after the Compton up-scattering is determined by the initial electron's spectrum shape. For the observation, the positive correlation of gamma-ray and GeV fluxes, and the negative correlation of the MeV and GeV spectral indices ($\alpha_\gamma$ and $\alpha_{GeV}$) exhibit. The finally boosted spectrum can be fitted by a broken power-law plus a thermal component.}
\label{fig:CartoonSpectrumBoosted}
\end{figure}



\clearpage

\input{table_t90_fitting_parameters.tex}  

\input{table_lat_parameters.tex}

\input{table_thermal_flux.tex}

\input{table_gbm_photon_index.tex}

\end{document}

%% file: table_t90_fitting_parameters.tex
\begin{deluxetable}{cclccc}
\renewcommand\arraystretch{1.2}
\tabletypesize{\small}
\tablewidth{0pt}
\tablecaption{The statistics of fitting the prompt spectrum. Data is from Fermi-GBM, the duration is T90 of $0-116$~s. We have used the following abbreviations: PL (Power-law), CPL (cutoff power-law) and BB (blackbody). The cutoff power-law plus blackbody model gives the best fit. \label{tab:promptModel}}

\tablehead{
\colhead{Instrument} & \colhead{Time} & \colhead{Model} & \colhead{Log(Likelihood)}& \colhead{AIC} & \colhead{BIC} \\
&(s)&&&&
 }
\startdata
 \emph{Fermi}-GBM              &  0 - 116              & PL                & 6970.41                   & 13944.85       & 13953.38       \\
                      &       (T90)           & CPL            & 6081.35                   & 12168.75       & 12181.52       \\
           &                   & Band           & 6080.67                  & 12169.42       & 12186.43       \\
                    &                   & PL+BB          & 5862.73                   & 11733.55       & 11750.56       \\
                    &                   & \underline{CPL+BB}   & 5678.19                   & 11271.46       & 11292.70       \\
                    &                   & Band+BB        & 5804.79                  & 11621.74       & 11647.21       \\
\hline
\enddata
\vspace{0.5cm}
\end{deluxetable}

%% file: table_lat_parameters.tex
\begin{deluxetable}{cccccc}
\renewcommand\arraystretch{1.1}
\tabletypesize{\small}
\tablewidth{0pt}
\tablecaption{Fermi-LAT spectral parameters, including the starting time  T$_{\mathbf{s}}$ and the ending time T$_{\mathbf{e}}$ of each time bin. The test statistics (TS) and the likelihood value. The photon index of a power-law model. Energy flux and count flux are integrated in the energy band of $100$~MeV to $1$~GeV. The fluxes without uncertainty are the upper limit. \label{tab:lat}}
\tablehead{
\colhead{Time} & \colhead{TS} & \colhead{Likelihood} & \colhead{Energy Flux}& \colhead{Count Flux} & \colhead{Photon Index} \\
(s)     & {}  & {} & ($\text{erg}~\text{cm}^{-2} \text{s}^{-1}$) & ($ \text{cm}^{-2} \text{s}^{-1}$) & {}
 }
\startdata
0.00     $\sim$ 2.70      & 11  & 4.32       & $4.52 \times 10^{-7}$          & $1.10 \times 10^{-3}$          &    -   		       \\
2.70     $\sim$ 4.10      & 186 & 44.80      & $3.34\pm0.77 \times 10^{-6} $  & $1.30\pm0.31 \times 10^{-2}$   & $-3.56\pm0.48$     \\
4.10     $\sim$ 4.46      & 112 & 5.52       & $5.70\pm2.02 \times 10^{-6} $  & $1.85\pm0.70 \times 10^{-2}$   & $-2.80\pm0.55$     \\
4.46     $\sim$ 4.77      & 140 & 9.76       & $8.37\pm2.65 \times 10^{-6} $  & $2.38\pm0.82 \times 10^{-2}$   & $-2.40\pm0.39$     \\
4.77     $\sim$ 5.06      & 135 & 12.34      & $7.95\pm2.67 \times 10^{-6} $  & $2.00\pm0.75 \times 10^{-2}$   & $-2.07\pm0.34$     \\
5.06     $\sim$ 5.46      & 134 & 10.80      & $4.64\pm1.73 \times 10^{-6} $  & $9.94\pm4.23 \times 10^{-3}$   & $-1.69\pm0.27$     \\
5.46     $\sim$ 5.83      & 129 & 10.17      & $5.40\pm1.95 \times 10^{-6} $  & $1.27\pm0.52 \times 10^{-2}$   & $-1.92\pm0.32$     \\
5.83     $\sim$ 6.15      & 153 & 7.24       & $8.12\pm2.57 \times 10^{-6} $  & $2.27\pm0.79 \times 10^{-2}$   & $-2.35\pm0.38$     \\
6.15     $\sim$ 6.51      & 183 & 14.91      & $8.55\pm2.66 \times 10^{-6} $  & $2.14\pm0.76 \times 10^{-2}$   & $-2.06\pm0.30$     \\
6.51     $\sim$ 6.89      & 124 & 9.37       & $6.16\pm2.05 \times 10^{-6} $  & $1.76\pm0.64 \times 10^{-2}$   & $-2.42\pm0.42$     \\
6.89     $\sim$ 7.10      & 104 & 12.76      & $8.25\pm3.19 \times 10^{-6} $  & $1.91\pm0.84 \times 10^{-2}$   & $-1.88\pm0.33$     \\
7.10     $\sim$ 7.46      & 155 & 12.40      & $7.12\pm2.27 \times 10^{-6} $  & $1.79\pm0.64 \times 10^{-2}$   & $-2.08\pm0.32$     \\
7.46     $\sim$ 7.98      & 128 & 17.95      & $4.40\pm1.48 \times 10^{-6} $  & $1.08\pm0.41 \times 10^{-2}$   & $-2.01\pm0.32$     \\
7.98     $\sim$ 8.54      & 150 & 16.50      & $5.44\pm1.58 \times 10^{-6} $  & $1.86\pm0.57 \times 10^{-2}$   & $-2.98\pm0.48$     \\
8.54     $\sim$ 9.20      & 157 & 21.68      & $4.14\pm1.27 \times 10^{-6} $  & $9.74\pm3.39 \times 10^{-3}$   & $-1.91\pm0.27$     \\
9.20     $\sim$ 10.40     & 125 & 32.85      & $1.99\pm0.66 \times 10^{-6} $  & $4.38\pm1.64 \times 10^{-3}$   & $-1.76\pm0.25$     \\
10.40    $\sim$ 11.50     & 114 & 27.70      & $1.51\pm0.60 \times 10^{-6} $  & $3.00\pm1.34 \times 10^{-3}$   & $-1.51\pm0.23$     \\
11.50    $\sim$ 13.33     & 126 & 32.48      & $1.55\pm0.47 \times 10^{-6} $  & $3.94\pm1.33 \times 10^{-3}$   & $-2.10\pm0.31$     \\
13.33    $\sim$ 16.50     & 114 & 30.07      & $8.04\pm2.57 \times 10^{-7} $  & $1.99\pm0.71 \times 10^{-3}$   & $-2.03\pm0.31$     \\
16.50    $\sim$ 19.86     & 108 & 26.85      & $5.39\pm2.04 \times 10^{-7} $  & $1.13\pm0.49 \times 10^{-3}$   & $-1.64\pm0.26$     \\
19.86    $\sim$ 22.15     & 108 & 34.64      & $9.41\pm3.27 \times 10^{-7} $  & $2.09\pm0.82 \times 10^{-3}$   & $-1.77\pm0.27$     \\
22.15    $\sim$ 30.56     & 104 & 28.64      & $1.54\pm0.69 \times 10^{-7} $  & $2.87\pm1.45 \times 10^{-4}$   & $-1.37\pm0.23$     \\
30.56    $\sim$ 38.82     & 120 & 51.83      & $2.42\pm0.91 \times 10^{-7} $  & $4.80\pm2.07 \times 10^{-4}$   & $-1.51\pm0.22$     \\
38.82    $\sim$ 46.75     & 127 & 30.40      & $1.90\pm0.78 \times 10^{-7} $  & $3.60\pm1.67 \times 10^{-4}$   & $-1.40\pm0.22$     \\
46.75    $\sim$ 61.32     & 72  & 45.04      & $1.68\pm0.56 \times 10^{-7} $  & $4.50\pm1.67 \times 10^{-4}$   & $-2.24\pm0.38$     \\
61.32    $\sim$ 93.00     & 75  & 46.39      & $7.56\pm2.88 \times 10^{-8} $  & $1.60\pm0.70 \times 10^{-4}$   & $-1.67\pm0.27$     \\
93.00    $\sim$ 250.00    & 49  & 138.78     & $3.52\pm1.33 \times 10^{-8} $  & $8.09\pm3.54 \times 10^{-5}$   & $-1.86\pm0.29$     \\
8300.00  $\sim$ 11650.00  & 25  & 761.13     & $2.32\pm1.59 \times 10^{-10}$  & $5.03\pm4.08 \times 10^{-7}$   & $-1.72\pm0.40$     \\
19650.00 $\sim$ 34600.00  & 4   & 2124.66    & $2.11 \times 10^{-10}$   		 & $5.14 \times 10^{-7}$   		  &     -    		   \\
79000.00 $\sim$ 126000.00 & 2   & 4967.33    & $1.31 \times 10^{-10}$   		 & $3.19 \times 10^{-7}$   		  &  	-		       \\
\hline
\enddata
\vspace{0.5cm}
\end{deluxetable}

%% file: table_thermal_flux.tex
\begin{deluxetable}{cccccc}
\renewcommand\arraystretch{1.2}
\tabletypesize{\small}
\tablewidth{0pt}
\tablecaption{Spectral parameters of the time bins having thermal component. The temperature, the best fitted model, the thermal and total flux, and the ratio of thermal flux. We use the following abbreviations: CPL (cutoff power-law) and BB (blackbody). Flux is defined in the energy band of $1$~keV to $10$~MeV. For the bins of $\sim 3$~s to $\sim 4$~s, Band+BB offers a very close goodness of fitting as CPL+BB, for the global consistency, and considering the time-integrated spectrum is best fitted by CPL+BB, here we perform all the thermal analysis using CPL+BB. \label{tab:thermalflux}}
\tablehead{
\colhead{Time} & \colhead{Model} & \colhead{Temperature} & \colhead{Thermal Flux}& \colhead{Total Flux} & \colhead{Ratio} \\
(s)&&(keV)&(erg cm$^{-2}$ s$^{-1}$)&(erg cm$^{-2}$ s$^{-1}$)
 }
\startdata
0.70$\sim$1.58&CPL+BB&230.3$^{+18.6}_{-19.5}$&2.84$^{+2.32}_{-1.27}$$\times$10$^{-5}$&0.80$^{+0.29}_{-0.19}$$\times$10$^{-4}$&0.35$^{+0.32}_{-0.18}$\\
1.58$\sim$1.71&CPL+BB&155.8$^{+10.2}_{-10.5}$&4.17$^{+1.91}_{-1.42}$$\times$10$^{-5}$&1.51$^{+0.28}_{-0.26}$$\times$10$^{-4}$&0.28$^{+0.14}_{-0.10}$\\
2.45$\sim$2.64&CPL+BB&173.9$^{+13.1}_{-12.6}$&3.88$^{+1.89}_{-1.40}$$\times$10$^{-5}$&1.76$^{+0.26}_{-0.24}$$\times$10$^{-4}$&0.22$^{+0.11}_{-0.09}$\\
2.64$\sim$2.88&CPL+BB&197.4$^{+16.5}_{-16.4}$&4.07$^{+1.91}_{-1.34}$$\times$10$^{-5}$&1.17$^{+0.28}_{-0.28}$$\times$10$^{-4}$&0.35$^{+0.18}_{-0.14}$\\
2.88$\sim$3.09&CPL+BB&187.6$^{+21.5}_{-21.9}$&2.36$^{+1.83}_{-1.11}$$\times$10$^{-5}$&1.01$^{+0.38}_{-0.29}$$\times$10$^{-4}$&0.23$^{+0.20}_{-0.13}$\\
3.09$\sim$3.21&CPL+BB&162.1$^{+12.2}_{-7.6}$&4.93$^{+1.71}_{-1.58}$$\times$10$^{-5}$&1.95$^{+0.85}_{-0.57}$$\times$10$^{-4}$&0.25$^{+0.14}_{-0.11}$\\
3.21$\sim$3.60&CPL+BB&149.2$^{+7.4}_{-7.6}$&3.16$^{+1.03}_{-0.79}$$\times$10$^{-5}$&1.36$^{+0.47}_{-0.31}$$\times$10$^{-4}$&0.23$^{+0.11}_{-0.08}$\\
3.60$\sim$3.74&CPL+BB&151.0$^{+11.2}_{-11.1}$&3.36$^{+1.65}_{-1.28}$$\times$10$^{-5}$&1.23$^{+0.84}_{-0.49}$$\times$10$^{-4}$&0.27$^{+0.23}_{-0.15}$\\
3.74$\sim$3.96&CPL+BB&140.4$^{+4.6}_{-4.5}$&7.11$^{+1.13}_{-1.01}$$\times$10$^{-5}$&2.23$^{+0.81}_{-0.55}$$\times$10$^{-4}$&0.32$^{+0.13}_{-0.09}$\\
3.96$\sim$4.10&CPL+BB&108.2$^{+5.7}_{-7.0}$&3.52$^{+1.85}_{-1.83}$$\times$10$^{-5}$&1.79$^{+1.71}_{-0.79}$$\times$10$^{-4}$&0.20$^{+0.22}_{-0.13}$\\
4.10$\sim$4.44&CPL+BB&114.8$^{+7.0}_{-7.2}$&1.89$^{+0.80}_{-0.59}$$\times$10$^{-5}$&0.94$^{+0.13}_{-0.11}$$\times$10$^{-4}$&0.20$^{+0.09}_{-0.07}$\\
4.44$\sim$4.51&CPL+BB&96.6$^{+14.8}_{-13.2}$&1.22$^{+1.57}_{-0.74}$$\times$10$^{-5}$&0.55$^{+0.23}_{-0.13}$$\times$10$^{-4}$&0.22$^{+0.30}_{-0.15}$\\
4.51$\sim$4.77&CPL+BB&111.1$^{+5.1}_{-5.0}$&3.45$^{+1.02}_{-0.72}$$\times$10$^{-5}$&0.85$^{+0.22}_{-0.16}$$\times$10$^{-4}$&0.41$^{+0.16}_{-0.12}$\\
4.77$\sim$4.95&CPL+BB&89.9$^{+3.9}_{-4.0}$&3.21$^{+0.89}_{-0.76}$$\times$10$^{-5}$&0.97$^{+0.33}_{-0.23}$$\times$10$^{-4}$&0.33$^{+0.15}_{-0.11}$\\
4.95$\sim$5.45&CPL+BB&91.0$^{+2.7}_{-2.7}$&2.45$^{+0.45}_{-0.39}$$\times$10$^{-5}$&0.70$^{+0.14}_{-0.10}$$\times$10$^{-4}$&0.35$^{+0.09}_{-0.07}$\\
5.45$\sim$5.51&CPL+BB&81.5$^{+5.4}_{-4.9}$&3.18$^{+1.30}_{-1.22}$$\times$10$^{-5}$&0.80$^{+0.95}_{-0.32}$$\times$10$^{-4}$&0.40$^{+0.49}_{-0.22}$\\
5.51$\sim$5.69&CPL+BB&47.9$^{+1.4}_{-1.5}$&1.43$^{+0.27}_{-0.26}$$\times$10$^{-5}$&0.42$^{+0.21}_{-0.12}$$\times$10$^{-4}$&0.34$^{+0.18}_{-0.12}$\\
\hline
\enddata
\vspace{0.5cm}
\end{deluxetable}

%% file: table_gbm_photon_index.tex
\begin{deluxetable}{cccccc}
\renewcommand\arraystretch{1.2}
\tabletypesize{\small}
\tablewidth{0pt}
\tablecaption{The lower photon index of time-resolved analysis. The value of photon index is chosen from the best model between  cutoff power-law (CPL) and cutoff power-law plus blackbody (CPL+BB). \label{tab:gbm_photon_index}}
\tablehead{
\colhead{Time} & \colhead{Model} & \colhead{Photon Index} & \colhead{Time}& \colhead{Model} & \colhead{Photon Index} \\
(s)&&&(s)&&
 }
\startdata
-0.07$\sim$0.03&CPL&-1.00$^{+0.17}_{-0.18}$&4.95$\sim$5.45&CPL+BB&-1.01$^{+0.03}_{-0.03}$\\
0.03$\sim$0.14&CPL&-1.20$^{+0.10}_{-0.10}$&5.45$\sim$5.51&CPL+BB&-0.89$^{+0.09}_{-0.16}$\\
0.14$\sim$0.29&CPL&-1.00$^{+0.05}_{-0.05}$&5.51$\sim$5.69&CPL+BB&-1.12$^{+0.08}_{-0.09}$\\
0.29$\sim$0.41&CPL&-0.84$^{+0.04}_{-0.04}$&5.69$\sim$5.81&CPL&-0.75$^{+0.05}_{-0.05}$\\
0.41$\sim$0.55&CPL&-0.72$^{+0.03}_{-0.03}$&5.81$\sim$6.00&CPL&-1.32$^{+0.05}_{-0.05}$\\
0.55$\sim$0.70&CPL&-0.58$^{+0.03}_{-0.03}$&6.00$\sim$6.44&CPL&-1.60$^{+0.04}_{-0.04}$\\
0.70$\sim$1.58&CPL+BB&-0.52$^{+0.03}_{-0.03}$&6.44$\sim$6.87&CPL&-1.75$^{+0.07}_{-0.09}$\\
1.58$\sim$1.71&CPL+BB&-0.66$^{+0.06}_{-0.05}$&6.87$\sim$8.22&CPL&-1.83$^{+0.02}_{-0.02}$\\
1.71$\sim$1.80&CPL&-0.51$^{+0.03}_{-0.03}$&8.22$\sim$9.57&CPL&-1.85$^{+0.02}_{-0.02}$\\
1.80$\sim$1.93&CPL&-0.71$^{+0.04}_{-0.04}$&9.57$\sim$12.40&CPL&-1.86$^{+0.06}_{-0.06}$\\
1.93$\sim$2.14&CPL&-0.91$^{+0.04}_{-0.04}$&12.40$\sim$15.55&CPL&-1.93$^{+0.07}_{-0.07}$\\
2.14$\sim$2.41&CPL&-0.96$^{+0.04}_{-0.04}$&15.55$\sim$15.87&CPL&-1.44$^{+0.06}_{-0.06}$\\
2.41$\sim$2.45&CPL&-0.89$^{+0.06}_{-0.06}$&15.87$\sim$16.17&CPL&-1.37$^{+0.05}_{-0.05}$\\
2.45$\sim$2.64&CPL+BB&-0.47$^{+0.04}_{-0.04}$&16.17$\sim$16.93&CPL&-1.35$^{+0.03}_{-0.03}$\\
2.64$\sim$2.88&CPL+BB&-0.58$^{+0.05}_{-0.05}$&16.93$\sim$17.32&CPL&-1.57$^{+0.05}_{-0.05}$\\
2.88$\sim$3.09&CPL+BB&-0.65$^{+0.04}_{-0.04}$&17.32$\sim$17.72&CPL&-1.56$^{+0.06}_{-0.06}$\\
3.09$\sim$3.21&CPL+BB&-0.73$^{+0.05}_{-0.06}$&17.72$\sim$20.40&CPL&-1.64$^{+0.03}_{-0.03}$\\
3.21$\sim$3.60&CPL+BB&-0.62$^{+0.04}_{-0.04}$&20.40$\sim$21.70&CPL&-1.88$^{+0.07}_{-0.07}$\\
3.60$\sim$3.74&CPL+BB&-0.68$^{+0.09}_{-0.08}$&21.70$\sim$23.33&CPL&-2.16$^{+0.03}_{-0.03}$\\
3.74$\sim$3.96&CPL+BB&-0.68$^{+0.05}_{-0.05}$&23.33$\sim$26.53&CPL&-1.96$^{+0.09}_{-0.09}$\\
3.96$\sim$4.10&CPL+BB&-0.37$^{+0.11}_{-0.10}$&26.53$\sim$33.08&CPL&-1.88$^{+0.07}_{-0.07}$\\
4.10$\sim$4.44&CPL+BB&-0.60$^{+0.04}_{-0.04}$&33.08$\sim$47.33&CPL&-1.93$^{+0.03}_{-0.03}$\\
4.44$\sim$4.51&CPL+BB&-0.92$^{+0.11}_{-0.11}$&47.33$\sim$73.50&CPL&-1.80$^{+0.11}_{-0.12}$\\
4.51$\sim$4.77&CPL+BB&-1.02$^{+0.04}_{-0.04}$&73.50$\sim$115.65&CPL&-1.85$^{+0.04}_{-0.04}$\\
4.77$\sim$4.95&CPL+BB&-0.85$^{+0.05}_{-0.05}$&115.65$\sim$150.00&CPL&-1.93$^{+0.09}_{-0.09}$\\
\hline
\enddata
\vspace{0.5cm}
\end{deluxetable}

%% file: 190114C.bbl
\begin{thebibliography}{10}

\bibitem{1997Natur.387..878M}
M.~R. {Metzger}, {\it et~al.\/}, {\it Nature\/} {\bf 387}, 878 (1997).

\bibitem{1997Natur.387..783C}
E.~{Costa}, {\it et~al.\/}, {\it Nature\/} {\bf 387}, 783 (1997).

\bibitem{1997Natur.386..686V}
J.~{van Paradijs}, {\it et~al.\/}, {\it Nature\/} {\bf 386}, 686 (1997).

\bibitem{2004ApJ...611.1005G}
N.~{Gehrels}, {\it et~al.\/}, {\it The Astrophysical Journal\/} {\bf 611}, 1005
  (2004).

\bibitem{2009ApJ...702..791M}
C.~Meegan, {\it et~al.\/}, {\it The Astrophysical Journal\/} {\bf 702}, 791
  (2009).

\bibitem{1998Natur.395..670G}
T.~J. {Galama}, {\it et~al.\/}, {\it Nature\/} {\bf 395}, 670 (1998).

\bibitem{1998Natur.395..663K}
S.~R. {Kulkarni}, {\it et~al.\/}, {\it Nature\/} {\bf 395}, 663 (1998).

\bibitem{2010ApJ...709L.172R}
F.~{Ryde}, {\it et~al.\/}, {\it The Astrophysical Journal Letters\/} {\bf 709},
  L172 (2010).

\bibitem{2014Sci...343...48M}
A.~{Maselli}, {\it et~al.\/}, {\it Science\/} {\bf 343}, 48 (2014).

\bibitem{2014Sci...343...42A}
M.~{Ackermann}, {\it et~al.\/}, {\it Science\/} {\bf 343}, 42 (2014).

\bibitem{2014Sci...343...38V}
W.~T. {Vestrand}, {\it et~al.\/}, {\it Science\/} {\bf 343}, 38 (2014).

\bibitem{2014Sci...343...51P}
R.~{Preece}, {\it et~al.\/}, {\it Science\/} {\bf 343}, 51 (2014).

\bibitem{2013Natur.500..547T}
N.~R. {Tanvir}, {\it et~al.\/}, {\it Nature\/} {\bf 500}, 547 (2013).

\bibitem{2017ApJ...848L..13A}
B.~P. {Abbott}, {\it et~al.\/}, {\it The Astrophysical Journal Letters\/} {\bf
  848}, L13 (2017).

\bibitem{2018ApJS..236...26L}
L.~{Li}, {\it et~al.\/}, {\it The Astrophysical Journal Supplement Series\/}
  {\bf 236}, 26 (2018).

\bibitem{2005ApJ...625L..95R}
F.~{Ryde}, {\it The Astrophysical Journal Letters\/} {\bf 625}, L95 (2005).

\bibitem{2009ApJ...702.1211R}
F.~Ryde, A.~Pe{\textquoteright}er, {\it The Astrophysical Journal\/} {\bf 702},
  1211 (2009).

\bibitem{GCN23688}
J.~D. {Gropp}, {\it et~al.\/}, {\it GRB Coordinates Network\/} {\bf 23688}
  (2019).

\bibitem{2019GCN.23716....1X}
S.~{Xiao}, {\it et~al.\/}, {\it GRB Coordinates Network\/} {\bf 23716} (2019).

\bibitem{GCN23690}
N.~{Tyurina}, {\it et~al.\/}, {\it GRB Coordinates Network\/} {\bf 23690}
  (2019).

\bibitem{GCN23693}
V.~{Lipunov}, {\it et~al.\/}, {\it GRB Coordinates Network\/} {\bf 23693}
  (2019).

\bibitem{GCN23692}
A.~{de Ugarte Postigo}, D.~A. {Kann}, C.~C. {Thoene}, L.~{Izzo}, {\it GRB
  Coordinates Network\/} {\bf 23692} (2019).

\bibitem{GCN23695}
J.~{Selsing}, J.~P.~U. {Fynbo}, K.~E. {Heintz}, D.~{Watson}, N.~{Dyrbye}, {\it
  GRB Coordinates Network\/} {\bf 23695} (2019).

\bibitem{GCN23699}
L.~{Izzo}, A.~{Noschese}, L.~{D'Avino}, M.~{Mollica}, {\it GRB Coordinates
  Network\/} {\bf 23699} (2019).

\bibitem{GCN23702}
J.~{Bolmer}, P.~{Schady}, {\it GRB Coordinates Network\/} {\bf 23702} (2019).

\bibitem{GCN23708}
A.~J. {Castro-Tirado}, {\it et~al.\/}, {\it GRB Coordinates Network\/} {\bf
  23708} (2019).

\bibitem{GCN23709}
D.~{Kocevski}, {\it et~al.\/}, {\it GRB Coordinates Network\/} {\bf 23709}
  (2019).

\bibitem{GCN23701}
R.~{Mirzoyan}, {\it et~al.\/}, {\it GRB Coordinates Network\/} {\bf 23701}
  (2019).

\bibitem{2009ApJ...706L.138A}
A.~A. {Abdo}, {\it et~al.\/}, {\it The Astrophysical Journal Letters\/} {\bf
  706}, L138 (2009).

\bibitem{2010ApJ...718L..14S}
C.~A. {Swenson}, {\it et~al.\/}, {\it The Astrophysical Journal Letters\/} {\bf
  718}, L14 (2010).

\bibitem{2015ApJ...798...10R}
R.~{Ruffini}, {\it et~al.\/}, {\it The Astrophysical Journal\/} {\bf 798}, 10
  (2015).

\bibitem{Xu:2013f70}
D.~Xu, {\it et~al.\/}, {\it The Astrophysical Journal\/} {\bf 776}, 98 (2013).

\bibitem{Wang:201968a}
Y.~Wang, {\it et~al.\/}, {\it The Astrophysical Journal\/} {\bf 874}, 39
  (2019).

\bibitem{Ruffini2019}
R.~{Ruffini}, {\it et~al.\/}, {\it GRB Coordinates Network\/} {\bf 23715}
  (2019).

\bibitem{Kann2019}
D.~A. {Kann}, {\it et~al.\/}, {\it GRB Coordinates Network\/} {\bf 23710}
  (2019).

\bibitem{GCN23983}
A.~{Melandri}, {\it et~al.\/}, {\it GRB Coordinates Network\/} {\bf 23983}
  (2019).

\bibitem{1998ApJ...506L..23P}
R.~D. {Preece}, {\it et~al.\/}, {\it The Astrophysical Journal Letters\/} {\bf
  506}, L23 (1998).

\bibitem{1998ApJ...497L..17S}
R.~{Sari}, T.~{Piran}, R.~{Narayan}, {\it The Astrophysical Journal Letters\/}
  {\bf 497}, L17 (1998).

\bibitem{2011MNRAS.415.3693R}
F.~{Ryde}, {\it et~al.\/}, {\it Monthly Notices of the Royal Astronomical
  Society\/} {\bf 415}, 3693 (2011).

\bibitem{2011MNRAS.410.2556D}
A.~{Dom{\'{\i}}nguez}, {\it et~al.\/}, {\it Monthly Notices of the Royal
  Astronomical Society\/} {\bf 410}, 2556 (2011).

\bibitem{2011ApJ...732...29C}
S.~B. {Cenko}, {\it et~al.\/}, {\it The Astrophysical Journal\/} {\bf 732}, 29
  (2011).

\bibitem{2013ApJS..209...11A}
M.~{Ackermann}, {\it et~al.\/}, {\it The Astrophysical Journals\/} {\bf 209},
  11 (2013).

\bibitem{2011ApJ...730..141Z}
B.-B. {Zhang}, {\it et~al.\/}, {\it The Astrophysical Journal\/} {\bf 730}, 141
  (2011).

\bibitem{zhang_2018}
B.~Zhang, {\it The Physics of Gamma-Ray Bursts\/} (Cambridge University Press,
  2018).

\bibitem{2014ApJ...784L..43B}
J.~M. {Burgess}, {\it et~al.\/}, {\it The Astrophysical Journal Letters\/} {\bf
  784}, L43 (2014).

\bibitem{2011ApJ...729..114A}
M.~{Ackermann}, {\it et~al.\/}, {\it The Astrophysical Journal\/} {\bf 729},
  114 (2011).

\bibitem{1988ApJ...333..646E}
R.~A. {Edelson}, J.~H. {Krolik}, {\it The Astrophysical Journal\/} {\bf 333},
  646 (1988).

\bibitem{Jiang:201586a}
Y.~G. Jiang, {\it et~al.\/}, {\it Monthly Notices of the Royal Astronomical
  Society\/} {\bf 456}, 3386 (2016).

\bibitem{Asano:2009eb8}
K.~Asano, S.~Guiriec, P.~Mészáros, {\it The Astrophysical Journal Letters\/}
  {\bf 705}, L191 (2009).

\bibitem{Zhang:2010564}
B.~Zhang, H.~Yan, {\it The Astrophysical Journal\/} {\bf 726}, 90 (2010).

\bibitem{Bosnjak:2012705}
Å.~BoÅ¡njak, P.~Kumar, {\it Monthly Notices of the Royal Astronomical Society:
  Letters\/} {\bf 421}, L39 (2012).

\bibitem{2010PhRvD..82d3002A}
F.~A. {Aharonian}, S.~R. {Kelner}, A.~Y. {Prosekin}, {\it Physical Review D\/}
  {\bf 82}, 043002 (2010).

\bibitem{Fan:20132be}
Y.-Z. Fan, {\it et~al.\/}, {\it The Astrophysical Journal\/} {\bf 776}, 95
  (2013).

\bibitem{2018arXiv181101839R}
R.~{Ruffini}, {\it et~al.\/}, {\it Journal of Cosmology and Astroparticle
  Physics, Submitted.\/} p. arXiv:1811.01839 (2018).

\bibitem{2017ApJ...834L..13W}
J.-J. {Wei}, B.-B. {Zhang}, L.~{Shao}, X.-F. {Wu}, P.~{M{\'e}sz{\'a}ros}, {\it
  The Astrophysical Journal Letters\/} {\bf 834}, L13 (2017).

\bibitem{Fynbo:2014e82}
J.~P.~U. Fynbo, {\it Science\/} {\bf 343}, 34 (2014).

\bibitem{2015ApJ...805...13L}
L.~{Li}, {\it et~al.\/}, {\it The Astrophysical Journal\/} {\bf 805}, 13
  (2015).

\bibitem{2015arXiv150708343V}
G.~Vianello, {\it et~al.\/}, {\it The 34th International Cosmic Ray
  Conference\/} p. arXiv:1507.08343 (2015).

\bibitem{1979ApJ...228..939C}
W.~Cash, {\it The Astrophysical Journal\/} {\bf 228}, 939 (1979).

\bibitem{2006ApJS..166..298K}
Y.~Kaneko, {\it et~al.\/}, {\it The Astrophysical Journal Supplement Series\/}
  {\bf 166}, 298 (2006).

\bibitem{2018arXiv181003129L}
L.~{Li}, {\it The Astrophysical Journal Supplement Series, Submitted\/}
  (2018).

\bibitem{2013ApJ...764..167S}
J.~D. Scargle, J.~P. Norris, B.~Jackson, J.~Chiang, {\it The Astrophysical
  Journal\/} {\bf 764}, 167 (2013).

\bibitem{Evans:2007iz}
P.~A. Evans, {\it et~al.\/}, {\it Astronomy and Astrophysics\/} {\bf 469}, 379
  (2007).

\bibitem{Evans:2009kx}
P.~A. Evans, {\it et~al.\/}, {\it Monthly Notices of the Royal Astronomical
  Society\/} {\bf 397}, 1177 (2009).

\bibitem{2001AJ....121.2879B}
J.~S. {Bloom}, D.~A. {Frail}, R.~{Sari}, {\it Astronomical Journal\/} {\bf
  121}, 2879 (2001).

\bibitem{2018ApJ...852...53R}
R.~{Ruffini}, {\it et~al.\/}, {\it The Astrophysical Journal\/} {\bf 852}, 53
  (2018).

\bibitem{1993ApJ...413..281B}
D.~{Band}, {\it et~al.\/}, {\it The Astrophysical Journal\/} {\bf 413}, 281
  (1993).

\bibitem{10.2307/2291091}
R.~E. Kass, A.~E. Raftery, {\it Journal of the American Statistical
  Association\/} {\bf 90}, 773 (1995).

\bibitem{10.2307/1912557}
Q.~H. Vuong, {\it Econometrica\/} {\bf 57}, 307 (1989).

\bibitem{10.2307/2289282}
W.~Cleveland, S.~Devlin, {\it Journal of the American Statistical
  Association\/} {\bf 83}, 596 (1988).

\bibitem{1998Natur.393..763A}
G.~{Amelino-Camelia}, J.~{Ellis}, N.~E. {Mavromatos}, D.~V. {Nanopoulos},
  S.~{Sarkar}, {\it Nature\/} {\bf 393}, 763 (1998).

\bibitem{2008JCAP...01..031J}
U.~{Jacob}, T.~{Piran}, {\it Journal of Cosmology and Astroparticle Physics\/}
  {\bf 1}, 031 (2008).

\bibitem{2009Sci...323.1688A}
A.~A. {Abdo}, {\it et~al.\/}, {\it Science\/} {\bf 323}, 1688 (2009).

\bibitem{2009Natur.462..331A}
A.~A. {Abdo}, {\it et~al.\/}, {\it Nature\/} {\bf 462}, 331 (2009).

\bibitem{Piran:1993790}
T.~Piran, A.~Shemi, R.~Narayan, {\it Monthly Notices of the Royal Astronomical
  Society\/} {\bf 263}, 861 (1993).

\bibitem{Drenkhahn:2002279}
G.~Drenkhahn, {\it Astronomy \& Astrophysics\/} {\bf 387}, 714 (2002).

\bibitem{Granot:2011142}
J.~Granot, S.~S. Komissarov, A.~Spitkovsky, {\it Monthly Notices of the Royal
  Astronomical Society\/} {\bf 411}, 1323 (2011).

\bibitem{Meszaros:19948f1}
P.~Mészáros, M.~J. Rees, {\it Monthly Notices of the Royal Astronomical
  Society\/} {\bf 269}, L41 (1994).

\bibitem{Toma:20090f0}
K.~Toma, X.-F. Wu, P.~Mészáros, {\it The Astrophysical Journal\/} {\bf 707},
  1404 (2009).

\bibitem{Ioka:2010f24}
K.~Ioka, {\it Progress of Theoretical Physics\/} {\bf 124}, 667 (2010).

\end{thebibliography}
